\documentclass[10pt, conference, compsocconf]{template/IEEEtran}

\usepackage[table]{xcolor}
\usepackage{cite}

\ifCLASSINFOpdf
\else
\fi
\usepackage[cmex10]{amsmath}
\usepackage{algorithm}
\usepackage{algorithmic}

\usepackage{fixltx2e}
\usepackage{ctable} 
\usepackage{multirow}

\usepackage[font=footnotesize,caption=false]{subfig}
\usepackage{float}

\usepackage{epsfig}
\usepackage{enumerate}

\newtheorem{definition}{Definition}

\newtheorem{condition}{Condition}

\renewcommand{\IEEEQED}{\IEEEQEDopen}
\newcommand\Tstrut{\rule{0pt}{2.2ex}}       
\newcommand\Bstrut{\rule[-0.9ex]{0pt}{0pt}} 
\newcommand{\TBstrut}{\Tstrut\Bstrut} 

\usepackage{amssymb}
\usepackage{mathtools}

\usepackage{booktabs,fixltx2e}
\usepackage[flushleft]{threeparttable}
\usepackage{url}

\usepackage[]{color}

\newcommand{\lowerBound}[1]{\underline{#1}}
\newcommand{\required}[1]{\ensuremath{\hat{#1}}}

\newcommand{\integerNumberSet}{\ensuremath{\mathbb{Z}}}
\newcommand{\rationalNumberSet}{\ensuremath{\mathbb{R}}}
\newcommand{\frameSizeSet}{\ensuremath{F}}

\newcommand{\numClients}{\ensuremath{n}}
\newcommand{\lr}{\ensuremath{\mathcal{LR}}}
\newcommand{\schedule}{\ensuremath{S}}
\newcommand{\scheduleEl}{\ensuremath{s}}

\newcommand{\frameSize}{\ensuremath{\textit{f}}}

\newcommand{\latency}{\ensuremath{\Theta}}
\newcommand{\requiredLatency}{\ensuremath{\required{\latency}}}

\newcommand{\bandwidth}{\ensuremath{\rho}}
\newcommand{\requiredBandwidth}{\ensuremath{\required{\bandwidth}}}

\newcommand{\scheduleMod}{\ensuremath{x}}
\newcommand{\minNumberDoneWork}{\ensuremath{r}}
\newcommand{\numberOfSlotsForOneProc}{\ensuremath{\phi}}
\newcommand{\MinNumberOfSlots}{\underline{\numberOfSlotsForOneProc}}

\newcommand{\arrivalTime}{\ensuremath{\textit{arr}}}
\newcommand{\finishingTime}{\ensuremath{\textit{fin}}}

\newcommand{\Criterium}{\ensuremath{\Phi}}
\newcommand{\assignment}{\ensuremath{\omega}}
\newcommand{\scheduleBP}{\ensuremath{a}}
\newcommand{\y}{\ensuremath{y}}
\newcommand{\belongs}{\ensuremath{b}}
\newcommand{\columnSet}{\ensuremath{\Omega}}
\newcommand{\dualsOne}{\ensuremath{\lambda}}
\newcommand{\dualsTwo}{\ensuremath{\sigma}}

\newcommand{\critMM}{\ensuremath{\Phi^{MM}}}
\newcommand{\critSM}{\ensuremath{\Xi^{sub}}}
\newcommand{\UBOnCriterion}{\ensuremath{\Phi^{UB}}}
\newcommand{\LBOnCriterion}{\ensuremath{\Phi^{LB}}}
\newcommand{\EstimatedLBOnCriterion}{\ensuremath{\overline{\Phi}^{LB}}}
\newcommand{\pattern}{\ensuremath{p}}
\newcommand{\patternIndex}{\ensuremath{k}}

\newcommand{\clientSet}{\ensuremath{C}}
\newcommand{\client}{\ensuremath{c}}

\newcommand{\slots}{\ensuremath{\phi}}

\newcommand{\rateParam}{\ensuremath{\beta}} 
\newcommand{\latencyParam}{\ensuremath{\gamma}} 
\newcommand{\size}{\ensuremath{sz}} 

\renewcommand{\baselinestretch}{0.91}

\newcommand{\beforeSpace}{1 mm}
\newcommand{\afterSpace}{1 mm}

\begin{document}
%
\title{Scalable and Efficient Configuration of Time-Division Multiplexed Resources}

\author{\IEEEauthorblockN{Anna Minaeva$^{1}$, P{\v r}emysl {\v S}{\r u}cha$^{1}$, Benny Akesson$^{1,2}$, Zden{\v e}k Hanz{\' a}lek$^{1}$}
\IEEEauthorblockA{$^{1}$Faculty of Electrical Engineering and Czech Institute of Informatics, Robotics and Cybernetics, Czech Technical University in Prague}
\IEEEauthorblockA{$^{2}$CISTER/INESC TEC and ISEP}
\IEEEauthorblockA{e-mail: \{minaeann, suchap, kessoben, hanzalek\}@fel.cvut.cz}
}


%

\newcommand{\mynote}[1]{\footnotemark{}\marginpar{\tiny\thefootnote[\bf #1]}}
\DeclarePairedDelimiter\ceil{\lceil}{\rceil}
\DeclarePairedDelimiter\floor{\lfloor}{\rfloor}

\maketitle

\begin{abstract}
\boldmath
Consumer-electronics systems are becoming increasingly complex as the
number of integrated applications is growing.  Some of these
applications have real-time requirements, while other non-real-time
applications only require good average performance.  For
cost-efficient design, contemporary platforms feature an increasing
number of cores that share resources, such as memories and interconnects.  
However, resource sharing causes contention 
that must be resolved by a resource arbiter, such as 
Time-Division Multiplexing.
A key challenge is to configure this arbiter to satisfy the bandwidth
and latency requirements of the real-time applications, while maximizing
the slack capacity to improve performance of their non-real-time
counterparts.  As this configuration problem is 
NP-hard, a sophisticated automated configuration method is required to avoid
negatively impacting design time. 

The main contributions of this article are: 1) An optimal
approach that takes an existing integer linear programming (ILP) model
addressing the problem and wraps it in a
branch-and-price framework to improve scalability. 2) A faster
heuristic algorithm that typically provides near-optimal solutions.
3) An experimental evaluation that 
quantitatively compares the branch-and-price approach to the
previously formulated ILP model and the proposed heuristic
. 4) A case study of an HD video and graphics
processing system that demonstrates the practical applicability of the approach.

\vspace{1em}
\noindent
\emph{Cite as}: Anna Minaeva, P{\v r}emysl {\v S}{\r u}cha, Benny Akesson, and Zden{\v e}k Hanz{\' a}lek, Scalable and Efficient Configuration of Time-Division Multiplexed Resources, \emph{Journal of Systems and Software}, Volume 113, March 2016, Pages 44-58, 0164-1212,
\url{https://doi.org/10.1016/j.jss.2015.11.019}.

\vspace{1em}
\noindent
\emph{Source code}: 
\url{https://github.com/CTU-IIG/BandP_TDM}

\end{abstract}


\vspace{3mm}

%
\IEEEpeerreviewmaketitle

\footnotetext{\copyright 2017. This manuscript version is made available under the CC-BY-NC-ND 4.0 license \url{http://creativecommons.org/licenses/by-nc-nd/4.0/}.
\\
This article is published in Journal of Systems and Software.
}

\section{Introduction}
\label{introduction}

\begin{figure}[ht]
\centering
\epsfig{file=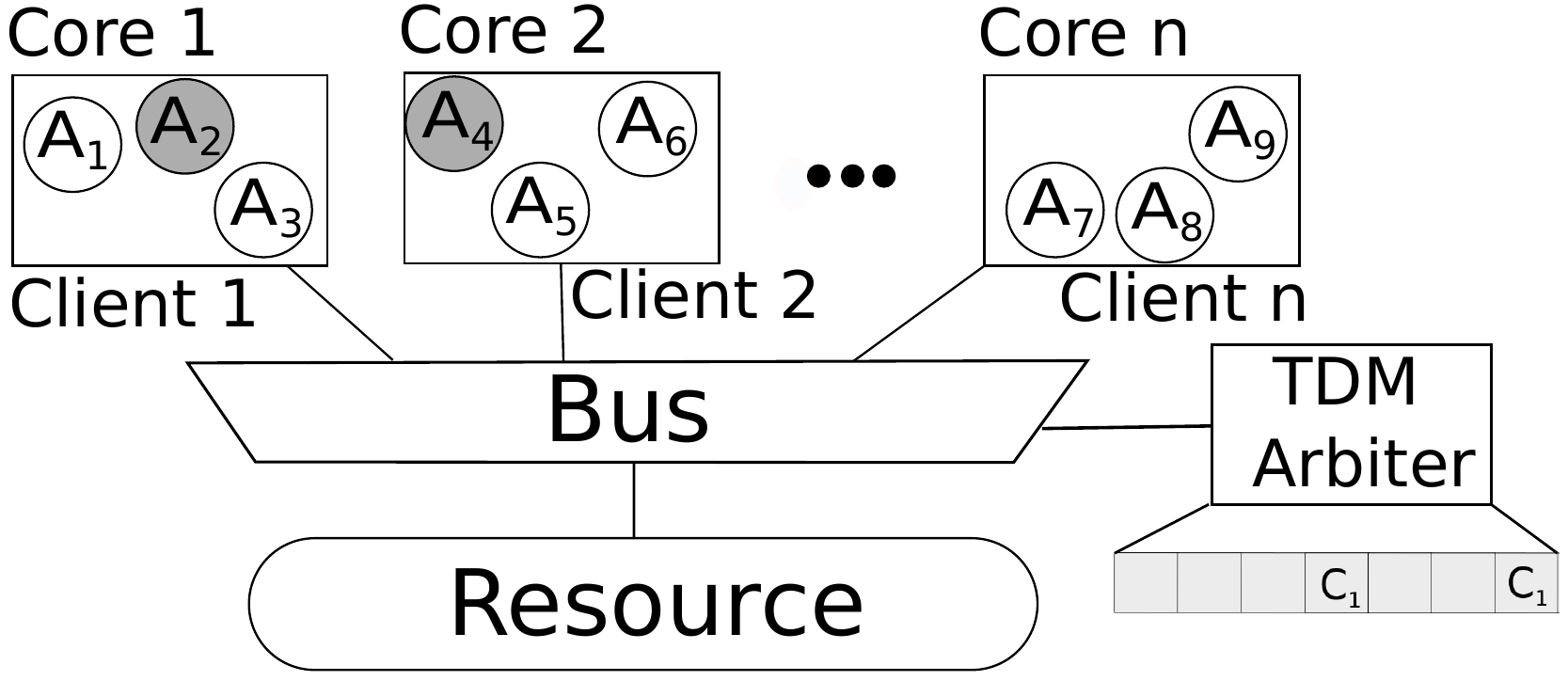,width=0.9\columnwidth}
\caption{Example of a multi-core system, where $\numClients$ applications ($A_l$)
with different real-time requirements are mapped to the cores. The cores act as resource clients ($c_i$) accessing a shared resource through an interconnect controlled by a TDM arbiter.}
\label{fig:problem_description}
\end{figure}

The trend of consumer-electronics systems becoming more and more
complex is not possible to overlook. An increasing number of
applications has resulted in a transition to multi-core platforms,
where the number of cores and hardware accelerators has been growing
exponentially; a trend that is expected to continue in the coming
decade~\cite{ITRS-2011}. The applications have different types of
requirements, as illustrated in Figure~\ref{fig:problem_description}. Some of them (colored white in
the figure) have \emph{real-time requirements} and must always
satisfy their deadlines, while other non-real-time applications
(colored gray) only require sufficient average performance~\cite{vanderWolf11}.  
The cores and accelerators access shared resources, such as memories, interconnects and
peripherals~\cite{Kollig09,Berkel09} on behalf of the applications
they execute and are referred to as resource \emph{clients}. However,
this resource sharing causes \emph{contention} between clients that
must be resolved by an \emph{arbiter}. Time-Division Multiplexing
(TDM) is a commonly used arbiter that schedules clients based on a
statically computed schedule with a fixed number of time slots.

An important challenge with TDM arbitration in these systems is to
find a schedule that assigns the time slots to the clients in a way that
satisfies the \emph{bandwidth and latency requirements} of the
real-time clients, while \emph{minimizing their resource utilization}
(maximizing slack capacity) to improve performance of the
non-real-time clients. This configuration process is required to
complete in reasonable time, even for large problems, to avoid
negatively impacting the design time of the systems, which is required
to stay unchanged~\cite{ITRS-2011}. This is particularly important since 
the schedule configuration may be a part of the design-space exploration,
thus requiring it to be repeated many times during system design.

The four main contributions of this article are: 1) We present an
exact approach that takes an existing integer linear programming (ILP)
model addressing the TDM configuration problem and
wraps it in a branch-and-price framework~\cite{Feilet10} to improve
its scalability. The computation time of this algorithm is optimized
using several techniques, including lazy constraints generation. 2) We
present a stand-alone heuristic algorithm that can be used to solve the
problem, providing a trade-off between computation
time and efficiency that is useful when sub-optimal solutions are
acceptable. 3) We experimentally demonstrate the improved scalability of the
branch-and-price approach and compare it both to the previously formulated
ILP model and to an existing heuristic. We also quantify the trade-off
between efficiency and computation time for the optimal and heuristic
algorithms. 4) We demonstrate the practical relevance of the approach
by applying it to a case study of an HD video and graphics
processing system. In addition, the source code of our approach is released as open-source software
and can be found in~\cite{BandP_TDM}.

The rest of this article is organized as follows. Related work is
discussed in Section~\ref{related_work}. Section~\ref{background}
proceeds by presenting background information necessary to understand
the main contributions of the article. Then, the
configuration problem is formalized in
Section~\ref{problem_formulation}, followed by a description of the existing ILP formulation in
Section~\ref{model}. The branch-and-price approach is introduced in Section~\ref{B&P}
and its computation time optimizations are discussed in Section~\ref{BP_details}.
The heuristic algorithm for slot assignment is then explained in
Section~\ref{heuristic}. Section~\ref{experiments} presents the experimental evaluation, before 
we demonstrate the practical applicability of our approach in a case study in Section~\ref{case_study}.
Lastly, the article is concluded in Section~\ref{conclusions}.

\section{Related work}
\label{related_work}

Scalability is a critical issue in system design, since design
time must remain unchanged despite an exponential increase in system 
complexity. Most works in the area of design automation that use exact optimization
techniques do not scale well enough to be able to manage the
complexity of future consumer electronics systems~\cite{Lukasie2012, gomony2015real, Yi2009,
  Lin2012,Hanzalek2010}, and only a few propose advanced techniques to address the
complexity problem. These techniques can be classified into two major
groups of approaches: 1) a decomposition of the problem into smaller
sub-problems, and 2) navigating the search smartly during design-space
exploration. The first approach deals with large problems by decomposing them
into many smaller problems. This method is used
in~\cite{Wildermann:2014:MDR,Liu2008}. The second branch of
improvements uses problem-specific information while searching the
design space, which is more efficient compared to using general
design-space exploration methods. Examples of this approach are shown
in~\cite{Reimann:2011:SSS:2024724.2024817} and~\cite{lukasiewycz12},
where the authors look for a minimal reason for constraint violation
and prevent this situation in the rest of the search, while using
boolean satisfiability and ILP approaches, respectively.

Another way of dealing with the scalability issue is to use a
heuristic approach. Some methodologies to configure TDM arbiters have
been proposed in the context of off-chip and on-chip networks. An
approach for synthesizing TDM schedules for TTEthernet with the goal
of satisfying deadlines for time-triggered traffic, while minimizing
the latency for rate-controlled traffic is proposed
in~\cite{tamas2012synthesis}.
The methodologies
in~\cite{Hansson07UNI,Lu_slotallocation} consider slot assignment in
contention-free TDM networks-on-chips. All of these approaches are heuristics
and the efficiencies of the proposed methods have not been quantitatively
compared to optimal solutions. Furthermore, the problem of scheduling
networks is different from ours, as it considers \emph{multiple
  resources} (network links) and is dependent on the problem of
determining paths through the network. 

The problem of TDM arbiter configuration with simplified client requirements 
is considered in~\cite{hassan2015framework}, where unlike this work, the authors propose a harmonic 
scheduling strategy. One of the two previous solutions to the problem 
considered in this article is the configuration methodology for multi-channel memory
controllers in~\cite{gomony2015real}. The authors apply a commonly used
heuristic for TDM slot assignment, called \emph{continuous allocation}~\cite{Goossens13CODES,Foroutan13DSD,Goossens13DATE,Vink08},
where slots allocated to a client appear consecutively in the
schedule. The reasons for its popularity are simplicity of
implementation and negligible computation time of the configuration
algorithm. However, with growing problem sizes, this strategy
results in significant over-allocation, making satisfaction of a
given set of requirements difficult. This is experimentally shown in
Section~\ref{experiments} when comparing to our approach.
The considered TDM configuration problem was furthermore addressed
in~\cite{Akesson15}, where an ILP formulation is proposed to solve the
problem. Although this solution shows good results
for systems of the past and present, complex future systems 
require a more scalable approach to avoid negatively impacting design time. 

Besides the difference in the problem formulation, this article
advances the state-of-the-art by being the \emph{first to apply a theoretically
well-founded advanced optimization approach, called
branch-and-price~\cite{Feilet10} in the field of consumer-electronics
systems design}. Branch-and-price combines both of the mentioned
approaches to manage complexity; it decomposes the problem into
smaller sub-problems and uses more sophisticated search-space
exploration methods. Although~\cite{schenkelaars2011optimal} applies 
branch-and-price to the problem of FlexRay scheduling in the 
automotive domain, this article gives more elaborate explanation of the approach
and concentrates on the computation time optimizations. Moreover, \emph{this work extends~\cite{Akesson15}
by using the previously proposed ILP model (Section~\ref{model}) as a building block in the
novel branch-and-price framework (Sections~\ref{B&P},~\ref{BP_details} and~\ref{heuristic}) to improve its scalability to satisfy the
needs of future design problems.}

\section{Background}
\label{background}

This section presents relevant background information to understand the work in this article. First, we present the concept of latency-rate servers, which is an abstraction of the service provided to a client by a resource arbiter. We then proceed by discussing how TDM arbitration fits with this abstraction and explain how to derive its latency and rate parameters.

\subsection{Latency-Rate Servers}
\label{lr_servers}

\begin{figure}
\centering
\epsfig{file=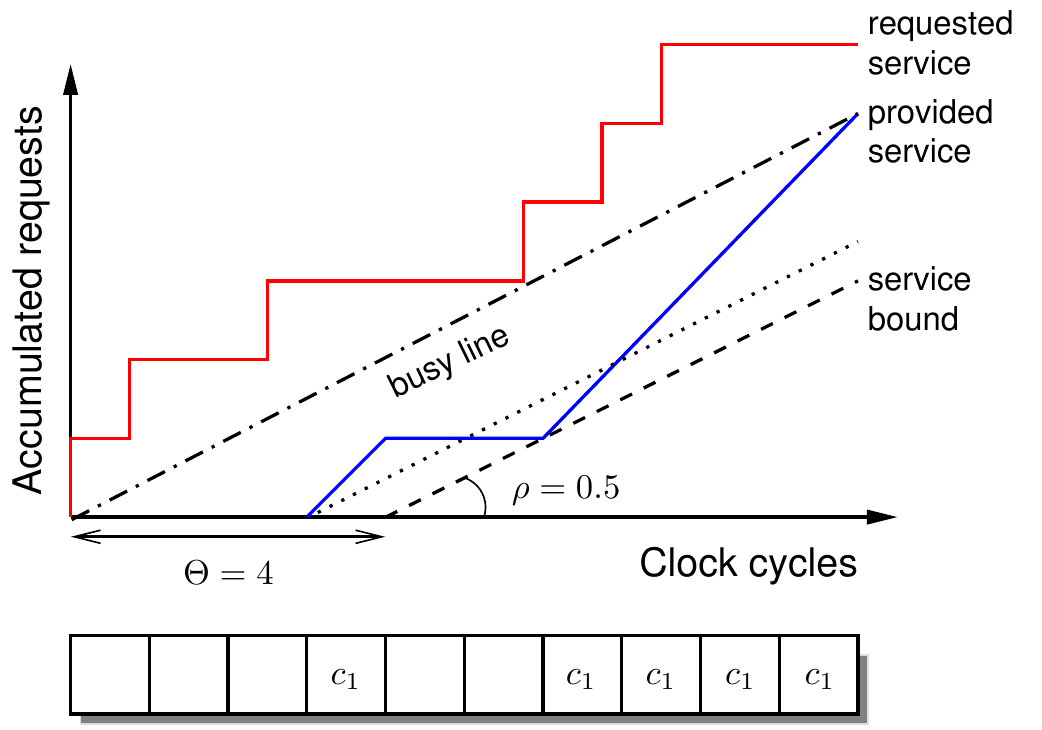,width=0.9\columnwidth}
\caption{A $\lr$ server and associated concepts for a client sharing a resource.}
\label{fig:latency_computation}
\end{figure}

Latency-rate ($\lr$)~\cite{Stiliadis98} servers is a shared resource abstraction that guarantees a client $\client_{i}$ sharing a resource a minimum allocated rate (bandwidth), $\rho_{i}$, after a maximum service latency (interference), $\latency_{i}$, as shown in Figure~\ref{fig:latency_computation}. The figure illustrates a client requesting service from a shared resource over time (upper solid red line) and the resource providing service (lower solid blue line). The $\lr$ service guarantee, the dashed line indicated as service bound in the figure, provides a lower bound on the amount of data that can be transferred to a client during any interval of time. 

The $\lr$ service guarantee is conditional and only applies if the client produces enough requests to keep the server busy. This is captured by the concept of \emph{busy periods}, which intuitively are periods in which a client requests at least as much service as it has been allocated ($\rho_{i}$) on average. This is illustrated in Figure~\ref{fig:latency_computation}, where the client is in a busy period when the requested service curve is above the dash-dotted reference line with slope $\rho_{i}$ that we informally refer to as the \emph{busy line}. We now have all the necessary ingredients to provide a formal definition of a $\lr$ server in Definition~\ref{def:lr_server}.

\vspace{\beforeSpace}
\begin{definition}[$\lr$ server]
\label{def:lr_server}
A $\lr$ server provides guarantees on minimum provided service $\minNumberDoneWork_{i}^{j}$ to a client $\client_{i}$ requesting the service during a busy period with duration $j$. These guarantees are expressed by Equation~\eqref{eq:lr_server} and are parametrized by service latency $\latency_{i}$ and rate $\rho_{i}$. The minimum non-negative constant $\latency_{i}$ satisfying Equation~\eqref{eq:lr_server} is the service latency of the server. 
\begin{equation}
\label{eq:lr_server}
\minNumberDoneWork_{i}^{j} \geq \max(0, \rho_{i} \cdot (j - \latency_{i}))
\end{equation}
\end{definition}
\vspace{\afterSpace}

The values of $\latency_{i}$ and $\rho_{i}$ of each client depend on the particular choice of arbiter and how it is configured. Examples of arbiters that belong to the class of $\lr$ servers are TDM, several varieties of the Round-Robin and Fair-Queuing algorithms~\cite{Stiliadis98}, as well as priority-based arbiters like Credit-Controlled Static-Priority~\cite{Akesson08RTCSA} and Frame-based Static Priority~\cite{akesson11-book}. The main benefit of the $\lr$ abstraction is that it enables performance analysis of systems with shared resources in a unified manner, irrespective of the chosen arbiter and configuration.
It has been shown in~\cite{Wiggers07SCOPES} that the worst-case finishing time of the $k^{\mathrm{th}}$ request from a client $\client_{i}$ in a $\lr$ server can be bounded 
according to Equation~\eqref{eq:finishing_time}, where $\size_{i}^{k}$ is the size of the request in number of required slots, $\arrivalTime_{i}^{k}$ is the arrival time and $\finishingTime_{i}^{k - 1}$ is the worst-case finishing time of the previous request from the client. 
This bound is visualized for the $k^{\mathrm{th}}$ request in Figure~\ref{fig:latency_computation}. 
Note that this bound is slightly pessimistic, but only serves to illustrate how the $\lr$ abstraction is used to compute the finishing time of requests. For optimized bounds and a quantitative evaluation of the abstraction, refer to~\cite{Shah13DATE}.

\begin{equation}
\label{eq:finishing_time}
\finishingTime_{i}^{k} =
\max(\arrivalTime_{i}^{k} + \latency_{i}, \finishingTime_{i}^{k - 1}) + \size_{i}^{k} / \rho_{i}
\end{equation}

Equation~\eqref{eq:finishing_time} forms the base for verification of applications at a higher level and does not make any assumptions on the applications by itself. Instead, restrictions on
the application are imposed by the higher level analysis frameworks.
For example, as shown in~\cite{Rodrigues15SOCP}, it is possible to integrate Equation~\eqref{eq:finishing_time} into a worst-case execution time estimation tool to enable bounds on execution time of an application sharing resources to be computed. However, these tools are often limited to analyzing applications executing on a single core. 
This restriction does not apply to verification based on the data-flow model of computation~\cite{Sriram00}, which can verify that distributed applications with data dependencies meet their real-time requirements, as demonstrated in~\cite{Nelson2015}. This type of verification is particularly suitable for throughput-oriented streaming applications, such as audio and video encoders/decoders~\cite{bhattacharyya1999synthesis,stuijk2008throughput,Vink08}, and wireless radios~\cite{Moreira07,Vink08}. In this case, Equation~\eqref{eq:finishing_time} is integrated into the data-flow graph as a data-flow component with two actors~\cite{Wiggers07SCOPES} before the analysis to capture the effects of resource sharing.

\subsection{Time-Division Multiplexing}
\label{tdm}

Having introduced $\lr$ servers as a general abstraction of shared resources, we proceed by showing how the abstraction applies to resources shared using TDM arbitration. We do this by first defining TDM arbitration and then show how the service latency and rate parameters of the corresponding $\lr$ server are derived.


A TDM arbiter operates by periodically repeating a schedule, or frame, with a fixed number of slots, $\frameSize$. The schedule comprises a number of slots, each corresponding to a single resource access with bounded execution time in clock cycles. Every client $\client_{i}$ is allocated a number of slots $\slots_{i}$ in the schedule at design time. 
The rate (bandwidth) allocated to a client, $\rho_{i}$, is determined purely by the number of allocated slots in the schedule and is
computed according to Equation~\eqref{eq:rho}. 

\begin{equation}
\label{eq:rho}
\rho_{i} = \slots_{i} / \frameSize
\end{equation}

The service latency, on the other hand, depends on the \emph{slot distribution} that determines how the allocated slots are distributed in the schedule. 

Although it may intuitively seem like computing the service latency is just a matter of identifying the largest gap between slots allocated to the client in the schedule, this is not correct. The reason is that according to Definition~\ref{def:lr_server}, the rate $\bandwidth$ has to be \emph{continuously} provided after the service latency $\latency$. The problem is illustrated in Figure~\ref{fig:latency_computation}, which shows a TDM schedule along with its corresponding service bound. As we can see, the service latency of $\latency=3$, which is the largest gap in the schedule, does not provide a conservative $\lr$ guarantee (it fails at time slot 6 in the TDM table) and the actual service latency of this schedule is $\latency=4$. This example shows us that the notion of service latency is more complex than it initially seems, making it harder to compute.
This article uses a very general way to compute the service latency, $\latency$, of a TDM arbiter by directly using Definition~\ref{def:lr_server}. 
The rate of a given schedule is known by Equation~\eqref{eq:rho} and the worst-case provided service to a client $\client_{i}$ during a busy period of any duration $j$ (i.e. $\minNumberDoneWork_{i}^{j}$) can be derived by analyzing the schedule (later shown in Section~\ref{model}). For a given schedule, it is hence only a matter of finding the minimum service latency
that satisfies the $\lr$ characterization in Equation~\eqref{eq:lr_server}.

In terms of implementation, we assume a scalable interconnect supporting TDM arbitration. A simple bus fails to scale as the number of clients are increasing, as the critical
path gets longer and prevents it from synthesizing at high frequencies. Although TDM-based networks-on-chips address this scalability problem, they behave like multiple TDM resources (one per link), resulting in a different configuration problem. Instead, we consider a pipelined tree-shaped interconnect supporting a distributed implementation of TDM arbitration, such as the memory tree proposed in~\cite{gomony2015generic}, which provides the required scalability yet behaves like a single resource. 
 

\section{Problem Formulation} 
\label{problem_formulation} 
The problem of finding a TDM slot allocation with a given frame size that satisfies the requirements of a \emph{set of clients}, while minimizing the rate allocated to real-time clients is formulated in this section. We refer to this problem as \emph{TDM Configuration Problem/Latency-Rate with given frame size} (TCP/LR-F).

An instance of the TCP/LR-F problem is defined by a tuple of requirements $\langle \clientSet, \requiredLatency, \requiredBandwidth, \frameSize \rangle$, where: 

\begin{itemize} 

\item{$\clientSet = \{\client_{1}, ..., \client_{\numClients}\}$ is the set 
of real-time clients that share a resource, where $\numClients$ is the number of 
clients.}

\item{$\requiredLatency \,= \, [\requiredLatency_{1}, \requiredLatency_{2}, ..., 
\requiredLatency_{\numClients}] \in 
\rationalNumberSet^{\numClients}_{\geq 0}$ and \mbox{$\requiredBandwidth = [\requiredBandwidth_{1}, 
\requiredBandwidth_{2}, ..., \requiredBandwidth_{\numClients}] \in 
\rationalNumberSet^{\numClients}_{\geq 0}$} are given 
\emph{service latency} (in number of TDM slots) and \emph{rate (bandwidth)} 
(required fraction of total available slots) \emph{requirements} of the clients, respectively.}
\item {$\frameSize$ is a given TDM frame size, $\frameSize \in 
\integerNumberSet^{+}$.}
\end{itemize} 


To satisfy the given requirements of a problem instance, we proceed by formalizing
a TDM schedule and its associated parameters:

\begin{itemize} 

\item{ The set $\frameSizeSet = 
\{1, 2, \cdots, \frameSize\}$ denotes TDM slots.} 

\item{$\schedule = [\scheduleEl_{1}, \scheduleEl_{2}, ..., 
\scheduleEl_{\frameSize}]$ is a schedule we want to find, where 
$\scheduleEl_{i} \in \{\clientSet \cup \varnothing\}$ indicates the client
scheduled in slot $i$ or $\varnothing$ (empty element) if the slot is not allocated.}

\item{$\slots = \{\slots_{1}, \slots_{2}, ..., \slots_{\numClients}\}$ is 
the number of slots allocated to each client, i.e. $\slots_{i} = 
\mid \{\scheduleEl_{j}\}: \scheduleEl_{j} = \client_{i} \mid$.}  

\item{$\latency = [\latency_{1}, \latency_{2}, ..., \latency_{\numClients}] 
\in \rationalNumberSet^{\numClients}_{\geq 0}$ and $\bandwidth = [\bandwidth_{1}, 
\bandwidth_{2}, ..., \bandwidth_{\numClients} ] \in 
\rationalNumberSet^{\numClients}_{\geq 0}$ are the \emph{service latency} and 
\emph{allocated rate}, respectively, provided by the TDM schedule.} 

\end{itemize} 

The goal of TCP/LR-F is to find a schedule $\schedule$
for $\numClients$ clients sharing the resource such that the objective function,
$\Criterium$, being the total allocated rate of all the real-time clients
in $\clientSet$ is minimized as shown in Equation~\eqref{eq:crit1}, while the service
latency and rate constraints~(Equations~\eqref{con:constr1}
and~\eqref{con:constr2} below) are fulfilled. This ensures that all
real-time requirements are satisfied while maximizing the unallocated
resource capacity available to non-real-time
clients, thus maximizing their performance.

\begin{equation}
\label{eq:crit1}
\textit{Minimize:}\: \sum_{\client_{i} \in \clientSet} \bandwidth_{i} = \Criterium 
\end{equation}

\begin{equation}
\label{con:constr1}
\bandwidth_i \geq \requiredBandwidth_{i}, \:\client_{i} \in \clientSet
\end{equation}

\begin{equation}
\label{con:constr2}
\latency_i \leq \requiredLatency_{i}, \:\client_{i} \in \clientSet
\end{equation}

Note that although the considered TCP/LR-F problem has a frame size $\frameSize$ as a given parameter, the problem of finding the best frame size is addressed in~\cite{Akesson15}, where both optimal and heuristic approaches are presented. It is also shown that the formulated problem with arbitrary frame size is NP-hard by transforming the Periodic Maintenance Scheduling Problem (PMSP)~\cite{bar2002minimizing} to the TCP/LR-F problem with arbitrary frame size. To prove NP-hardness of our problem with given frame size using the same logic, the frame size $\frameSize$ of the instance we transform PMSP to is set to the least common multiple of the clients periods, making it a special case that is covered by the existing proof. Note that this choice of frame size proves NP-hardness of the \mbox{TCP/LR-F} problem in general, i.e. not only for instances with this frame size.


\section{ILP Model}
\label{model}

Since the branch-and-price approach is built on the existing ILP formulation, it is briefly introduced first. 
For a more elaborate description of the formulation and its optimizations, refer to~\cite{Akesson15}.
In the ILP formulation, the schedule S is represented using binary decision variables $\scheduleMod_{i}^{j}$ indicating that slot $j \in \frameSizeSet$ is allocated to a client $\client_{i} \in \clientSet$. This is defined as:

\begin{equation*}
\scheduleMod_{i}^{j} =
  \begin{cases}
    1, & \text{if slot $j$ is allocated to client $\client_{i}$}.\\
    0, & \text{otherwise}.
  \end{cases}
\end{equation*}  

The minimization criterion~\eqref{eq:tmindCrit} is reformulated from Equation~\eqref{eq:crit1} in terms of the variables presented above. The solution space is defined by four constraints: Constraint~\eqref{con:tmind1} states that a slot can be allocated to maximally one client. Constraint~\eqref{con:tmind2} then dictates that enough slots must be allocated to a client to satisfy its rate requirement, which is computed according to Equation~\eqref{eq:rho}. The following two constraints focus on the worst-case provided service offered by the TDM schedule to a client, $\lowerBound{\minNumberDoneWork}^j_i$ ($\lowerBound{\minNumberDoneWork}^j_i \leq \minNumberDoneWork_i^j$), where $\minNumberDoneWork_i^j$ corresponds to the lower solid blue line labeled 'provided service' in Figure~\ref{fig:latency_computation}. Constraint~\eqref{con:tmind3} states that the worst-case provided service to a client $\client_{i}$ during a busy period of any duration $j$ starting in any slot $k$ cannot be larger than the service provided by its allocated slots.

Lastly, Constraint~\eqref{con:tmind4} states that the worst-case provided service of the client, $\lowerBound{\minNumberDoneWork}_{j}^{i}$, must satisfy its $\lr$ requirements and is a straight-forward implementation of Definition~\ref{def:lr_server}. 



\begin{equation}
\label{eq:tmindCrit}
\textit{Minimize:} \; \frac{\sum_{\client_{i} \in \clientSet}\sum_{j \in \frameSizeSet}\scheduleMod_{i}^{j}}{\frameSize}.
\end{equation} 

\emph{subject to}:
\begin{equation}
\label{con:tmind1}
\sum_{\client_{i} \in \clientSet} \scheduleMod_{i}^{j} \leq 1, \qquad j \in \frameSizeSet.
\end{equation}

\begin{equation}
\label{con:tmind2}
\sum_{j = 1}^{\frameSize} \scheduleMod_{i}^{j} \geq \frameSize \cdot  \hat{\rho}_{i}, \qquad \client_{i} \in \clientSet.
\end{equation}


\begin{equation}
\label{con:tmind3}
\lowerBound{\minNumberDoneWork}_{i}^{j} \leq \sum_{l = k}^{(k + j) \bmod \frameSize} \scheduleMod_{i}^{l},  \qquad k \in \frameSizeSet,\: \client_{i} \in \clientSet, \: j \in \frameSizeSet.
\end{equation}

\begin{equation}
\label{con:tmind4}
\lowerBound{\minNumberDoneWork}_{i}^{j} \geq \requiredBandwidth_{i} \cdot  (j - \requiredLatency_{i}), \qquad j \in \frameSizeSet, \: \client_{i} \in \clientSet.
\end{equation}


After introducing the basic ILP model of TCP/LR-F, we proceed by discussing a few computation time optimizations. The first optimization exploits that an \emph{increased lower bound on the number of slots allocated to a client}, $\MinNumberOfSlots_{i}$ can be found by considering both its rate (first part) and service latency (second part) requirements in Equation~\eqref{con:tmind7}. Unlike the rate requirement, the number of slots required to satisfy the service latency requirement depends on where the slots are allocated in the frame, which is not known beforehand. This lower bound is obtained by assuming an equidistant allocation, which results in the minimum number of slots required to be allocated to satisfy the service latency requirement. 

\begin{equation}
\label{con:tmind7}
\sum_{j \in \frameSizeSet} \scheduleMod_{i}^{j} \geq \MinNumberOfSlots_{i} = \max(\left\lceil \requiredBandwidth_{i} \cdot \frameSize \right\rceil, 
\left\lceil \frac{\frameSize}{\requiredLatency_{i} + 1} \right\rceil).
\end{equation} 

For example, having requirements $\requiredBandwidth_{1} =0.5$ and $\requiredLatency_{1} = 3$ for client $\client_{1}$ and a frame size $\frameSize = 10$, the lower bound on the number of allocated slots $\MinNumberOfSlots_{1}$ is the maximum 
of the $\left\lceil 0.5 \cdot 10 \right\rceil = 5$ slots required to satisfy the bandwidth requirement and the $\left\lceil \frac{10}{3 + 1} \right\rceil = 3$ slots required to allocate each fourth slot in the TDM table to the client. Thus, the lower bound for client $\client_{1}$ equals $ \MinNumberOfSlots_{1} = 5$, which in this case is determined by its bandwidth requirement.

The second optimization \emph{removes redundant constraints} generated by Constraints~\eqref{con:tmind3} and~\eqref{con:tmind4}.
As one can see, $\frameSize^{2} \cdot \numClients$ constraints are generated by Constraint~\eqref{con:tmind3} and $\frameSize \cdot \numClients$ constraints by Constraint~\eqref{con:tmind4}. However, it is not necessary to generate Constraints~\eqref{con:tmind3} and~\eqref{con:tmind4} for $j < \requiredLatency_{i}$, since the service bound provided by the $\lr$ guarantee is always zero in this interval by Definition~\ref{def:lr_server}. This is clearly seen in Figure~\ref{fig:latency_computation}, where the provided service curve is zero for the first four slots.

Additional constraints can be removed if more slots are required to satisfy the service latency requirements than the rate requirements, i.e. when the second term in the max-expression in Constraint~\eqref{con:tmind7} is dominant. In the remainder of this article, we refer to clients with this property as \emph{latency-dominated}, as opposed to \emph{bandwidth-dominated}, clients. As shown in~\cite{Akesson15}, for latency-dominated clients Constraints~\eqref{con:tmind3} and~\eqref{con:tmind4} only need to be generated for a single point where $j = \floor{\requiredLatency_i} + 1$. 

The third optimization reduces the solution space by \emph{reducing rotational symmetry}. This means that for any given TDM schedule, $\frameSize-1$ similar schedules can be generated by rotating the given schedule and wrapping around its end. The problem is that all these schedules have the same criterion value and only one of them needs to be in the considered solution space. Constraint~\eqref{con:tmind6} addresses this problem by adding a constraint that fixes the allocation of the first slot to the client with the smallest minimum number of required slots $\MinNumberOfSlots_{i}$ (defined in Constraint~\eqref{con:tmind7}). 
This particular choice of client has been experimentally determined to significantly reduce the computation time of the solver. 

\begin{equation}
\label{con:tmind6} 
\scheduleMod_{t}^{1} = 1, t = argmin_{\client_{i} \in \clientSet} \ \MinNumberOfSlots_{i}.
\end{equation}


\section{Branch-and-Price Approach}
\label{B&P}
The presented ILP model finds optimal solutions in reasonable time for current multi-core systems. However, despite the optimizations, it does not scale to many-core systems with 32 or more clients. To expand the range of problems that we are able to solve by the ILP model described in the previous section, a branch-and-price approach~\cite{Feilet10} is introduced, which uses the ILP problem formulation from the previous section as a building block. Branch-and-price allows solving instances of the TCP/LR-F problem with larger number of clients where the ILP formulation becomes too slow. The first reason for this behavior is that it does not need as many constraints in the ILP formulation for non-latency-dominated clients, where $\frameSize$~$^2 \cdot \numClients$ Constraints \eqref{con:tmind3} are required in the ILP. Moreover, it reduces the number of explored symmetrical solutions and typically has a smaller branching tree. Both of these properties result in a significantly reduced computation time for large problem instances compared to the previously described ILP model. The structure of this section is the following. First, background on the branch-and-price approach is provided. Then, all the necessary problem-dependent parts of the algorithm are described. The computation time optimizations of the algorithm are given in the following section. 

\subsection{Preliminaries}
\label{BP_background}
Branch-and-price is an exact method to solve ILP problems, which combines column generation and branch-and-bound approaches. In order to obtain a problem formulation for the branch-and-price approach, Dantzig-Wolfe decomposition~\cite{Feilet10} is performed on the ILP model from the previous section. At higher level, this decomposition transforms the space of binary variables $\scheduleMod_{i}^{j}$ of the ILP model into the space of complete solutions for individual clients, i.e. branch-and-price works with \emph{complete schedules for individual clients} instead of dealing with allocation of single slots. The solutions for a single client are called columns.

The process of applying Dantzig-Wolfe decomposition on the ILP model results in an ILP~\textit{master model} $MM(\columnSet)$ that contains a \emph{set of all possible columns} $\columnSet = \{\columnSet_1, \columnSet_2,\cdots,\columnSet_{\numClients}\}$.
Columns are iteratively generated by a so called~\textit{sub-model}, here an ILP model 
for a single client. Then, they are combined into a complete solution for all clients by the master model. Note that each client requires its own instance of the sub-model, since they have distinct requirements. 



\begin{figure}
\centering
\includegraphics[width=0.7\columnwidth]{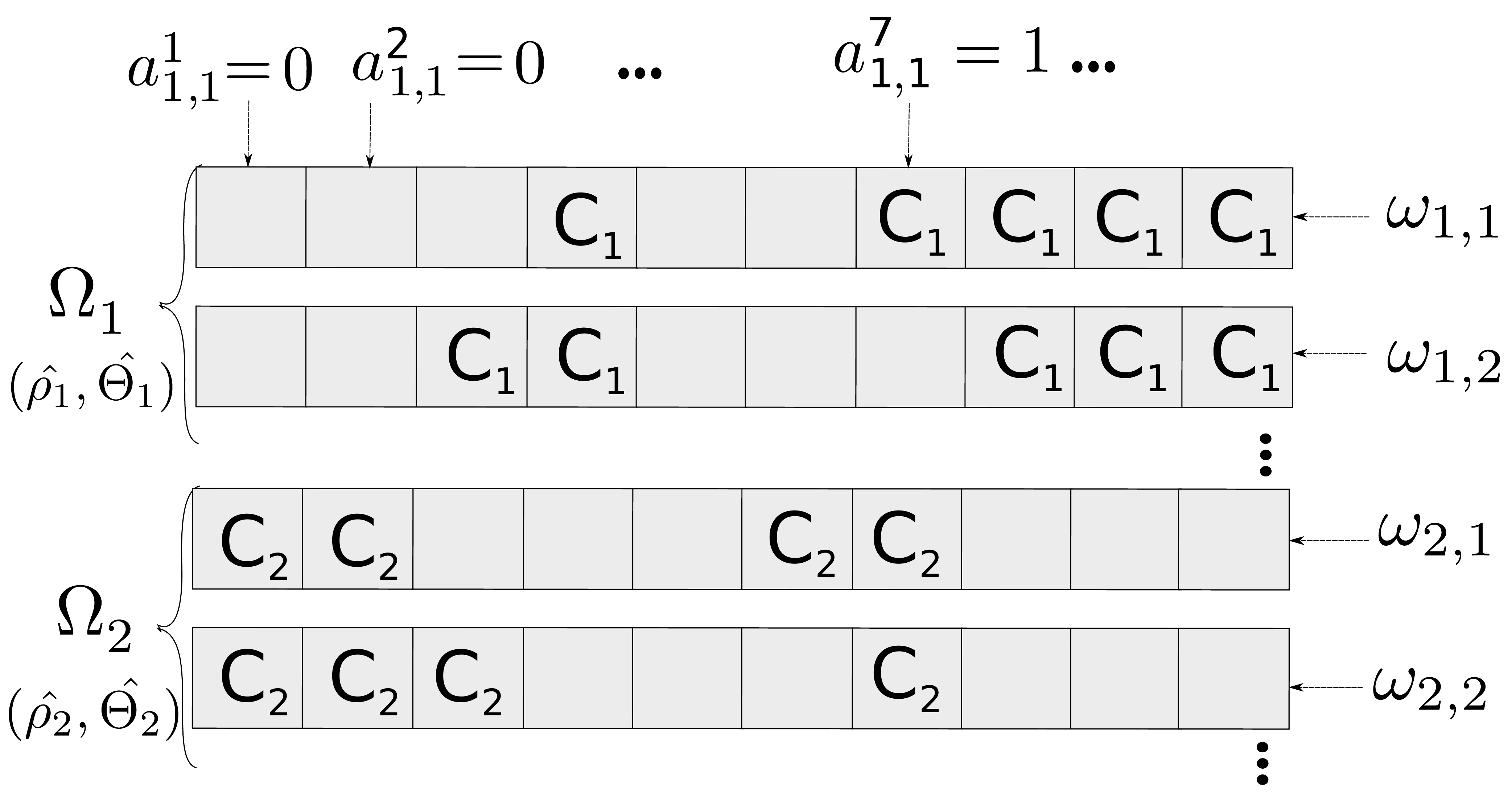}
\caption{Example of a set of columns for the restricted master model.}
\label{fig:column}
\end{figure}

An example column set is shown in Figure~\ref{fig:column}. In the considered TCP/LR-F problem, columns are complete TDM schedules for individual clients. 
Here, the set of columns $\columnSet_1$ for the first client with requirements $\requiredBandwidth_1, \requiredLatency_1$ contains two columns on the top of the figure and the set of columns $\columnSet_2$ for the second client is at the bottom. The decision variables $\assignment_{i,\patternIndex}$ indicate whether or not column $\pattern_{\patternIndex}$ is included in the schedule for client $\client_{i}$. One of the possible solutions here is to use column 2 for the first client and column 1 for the second one, i.e. $\assignment_{1,1} = 0$, $\assignment_{1,2} = 1$, $\assignment_{2,1} = 1$, $\assignment_{2,2} = 0$. Each column $\pattern_{\patternIndex}$ is defined via a set of coefficients $\scheduleBP_{i,\patternIndex}^j$. This coefficient is equal to $1$ if column $\pattern_{\patternIndex}$ allocates slot $j$ to client $\client_{i}$ and $0$ otherwise.

A drawback of using columns instead of the binary variables $\scheduleMod_{i}^{j}$ of the ILP model is the large number of possible columns. However, it is sufficient to gradually generate only the most promising ones and expand the search space of solutions step-by-step. At a certain (final) moment it can be proven (see~\cite{Feilet10}) that the optimal solution is found. A master model that considers only a subset of columns $\columnSet^R = \{\columnSet_1^R, \columnSet_2^R,\cdots,\columnSet_{\numClients}^R\} \subseteq \columnSet$ is called \textit{the Restricted Master Model} and is denoted as $MM(\columnSet^R)$. 


Thus, the idea, described above, is known as the \emph{column generation} approach. Since column generation is only able to solve the linear relaxation of the master model, it is necessary to extend this approach with a \emph{branch-and-bound} technique in order to get an integer solution. This combination is known in the literature as \textit{branch-and-price}.

\subsection{Outline of the algorithm}
\label{BP_outline}

\begin{figure}
\centering
\includegraphics[width=0.95\columnwidth]{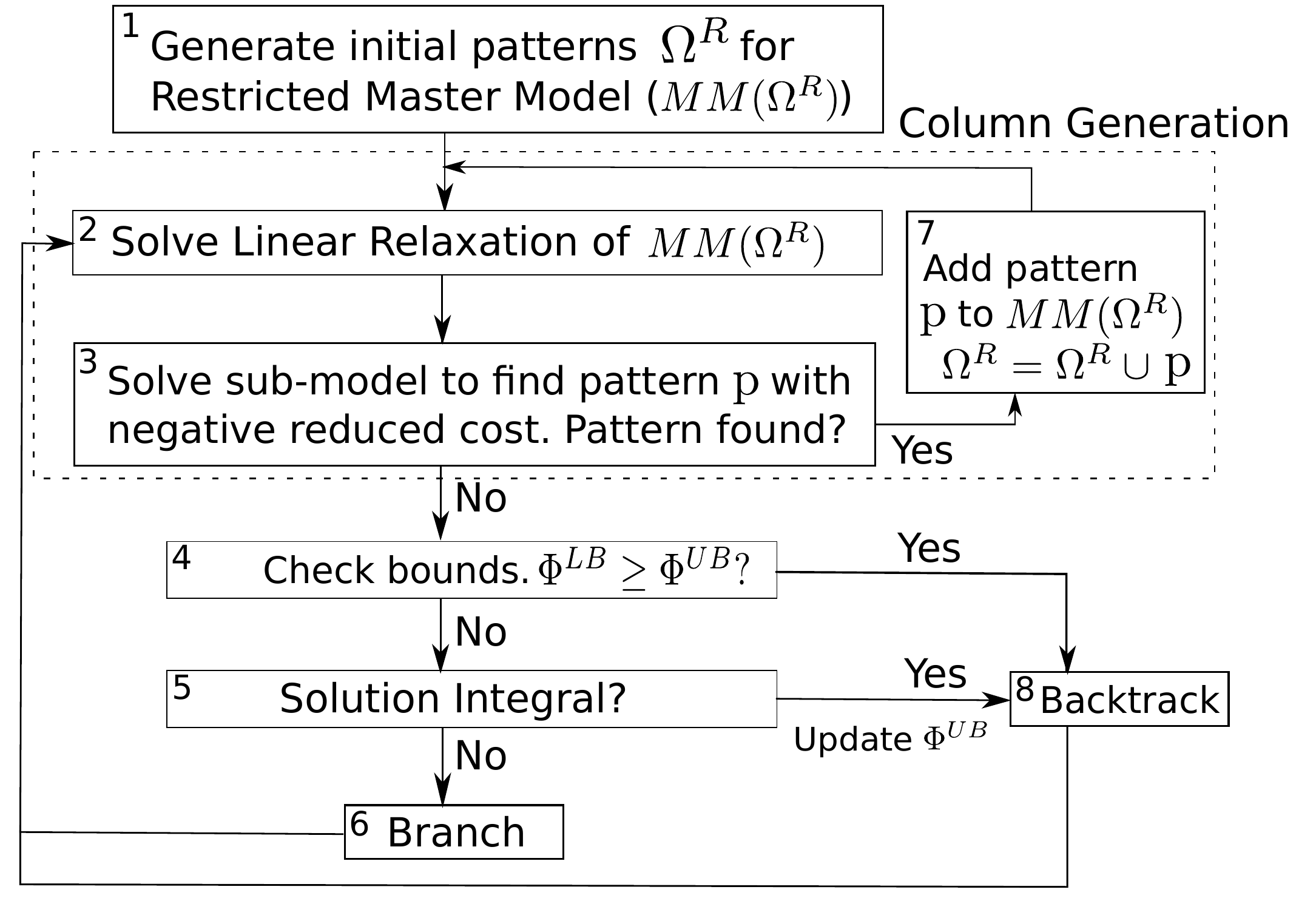}
\caption{Outline of the branch-and-price algorithm.}
\label{fig:BPScheme}
\end{figure}

The overall scheme of the branch-and-price algorithm in 8 steps is shown in Figure~\ref{fig:BPScheme}. 
First of all, the algorithm must generate a set of initial columns $\columnSet^R$ in Step~1 using a heuristic. Note that the quality of the initial columns only affects the computation time and not the optimality of the solution.
Step~2 starts the process of column generation by solving the linear relaxation of the restricted master model. The output of this step is quantitative directions (dual values) that guide the column search for the sub-model. Next, a new column for some client is constructed by the sub-model (Step 3) subject to the directions obtained in the previous step by $MM(\columnSet^R)$. If a new promising column for any client is found, the column is added to $\columnSet^R$ and the next iteration of the column generation algorithm starts (back to Step~2). Otherwise, the optimal linear solution of the relaxed $MM(\columnSet^R)$ is \emph{a lower bound for the optimal integral solution} and is denoted as $\LBOnCriterion$. If bounding takes place in Step~4, i.e. this branch is already worse or the same as the best solution known so far, the current node is closed in Step~8. In case the branch is still promising, Step~5 checks whether or not the solution obtained by column generation is integral. In case it is, a new candidate solution to the initial integer master model is found. This solution defines a new \emph{upper bound on the criterion}, $\UBOnCriterion$, which is updated before the node is closed in Step~8. In case neither bounding nor the check on integrality closes the node, branching takes place.

Using the branch-and-bound technique means having a branching tree, which is in essence a set of nodes with parent-child relationships. The node is defined by a partial solution, i.e. a chain of slot assignment decisions made in the parent nodes. A decision is to impose/forbid assignment of a slot to a client. At the beginning, the first slot is fixed in the root node, as mentioned in the optimizations of the ILP model in Section~\ref{model}.
The column generation procedure (Steps 2, 3, 7) together with bounding (Step~4) and checking solution on integrality (Step~5) are launched in each node. In case the node is not closed, a child node is generated with a new decision, i.e. which slot to impose/forbid assignment to which client. 



\subsection{Master Model Formulation}
\label{BP_MM}
The master model combines TDM tables for individual clients in order to provide a feasible solution, where each slot is allocated at most once in the schedule and requirements of all clients are satisfied. Moreover, it searches for the feasible schedule with the minimum total allocation. The master model is formulated as an integer linear programming problem, but in column generation it is solved as a linear problem, as stated earlier.

For TCP/LR-F, the Dantzig-Wolfe decomposition of the ILP model from Section~\ref{model} results in the following master model. The columns in $\columnSet^R$ are defined via coefficients $\scheduleBP_{i,\patternIndex}^j$, indicating whether or not a slot in a column is allocated to a client, defined according to
  $$ \scheduleBP_{i,\patternIndex}^j =\begin{cases}
    1, & \text{if slot $j$ is allocated to client $\client_{i}$ in $\pattern_{\patternIndex} \in \columnSet_i$}\\
    0, & \text{otherwise}.
  \end{cases} $$ 

Decision variables indicating whether client $\client_{i} \in \clientSet$ uses column $\pattern_{\patternIndex} \in \columnSet_i$, are defined as
 $$ \assignment_{i,\patternIndex} =\begin{cases}
    1, & \text{if client $\client_{i}$ uses column $\pattern_{\patternIndex} \in \columnSet_i$}\\
    0, & \text{otherwise}.
  \end{cases} $$ 
 It is a binary decision variable, which can be interpreted in the linear relaxed master model as the weight or ''probability'' of using column $\pattern_{\patternIndex}$ for client $\client_{i}$. 
 
Furthermore, another set of decision variables is introduced. Variables $\y_j$ reflect the over-allocation of slot $j$ in the final solution, i.e. $ \y_j =\max(0, v_j)$, where slot $j$ is allocated by $v_j - 1$ clients. It means that any feasible solution has $ \y_j =0, \forall j \in \frameSizeSet$. These variables are introduced in order to have an initial set of columns $\columnSet^R$ even in cases where a feasible solution could not be found in reasonable time in Step 1 of Figure~\ref{fig:BPScheme}. This is important since it is in general not possible to find $\columnSet^R$ that contains a feasible solution in polynomial time.

Minimization of the total allocated rate for the restricted master model is formalized in the objective function

\begin{equation}\label{BP1}
Minimize:\ \frac{\sum_{\client_{i} \in \clientSet}\sum_{\pattern_{\patternIndex} \in \columnSet_i^R} \slots_{i,\patternIndex}  \cdot \assignment_{i,\patternIndex}}{\frameSize} + M' \cdot \sum_{j \in \frameSizeSet} \y_j = \critMM,
\end{equation} 
where $\slots_{i,\patternIndex}$ is the total number of allocated slots to client $\client_{i}$ in column $\pattern_{\patternIndex}$ and $M'$ is some sufficiently big number, in this work chosen to be $M' = 10$. The sum of over-allocation variables $\y_j$ are multiplied by $M'$ to provide a major penalty for overlapping (infeasible) schedules to ensure they are weeded out quickly.

The master model comprises only two constraints. The first one, Constraint~\eqref{BP2}, is meant for counting the number of times slot $j$ is over-allocated in the schedule, which is expressed by variable $\y_j$.

\begin{equation} \label{BP2}
\sum_{\client_{i} \in \clientSet} \sum_{\pattern_{\patternIndex} \in \columnSet_i} \scheduleBP_{i,\patternIndex}^j \cdot \assignment_{i, \patternIndex} \leq \y_j + 1, \qquad j \in \frameSizeSet.
\end{equation}

Coefficients $\belongs_{i,\patternIndex}$ are introduced for the second constraint, indicating whether column $\pattern_{\patternIndex}$ was constructed for client $\client_{i}$:
$$ \belongs_{i,\patternIndex}=\begin{cases}
    1, & \text{if $\pattern_{\patternIndex} \in \columnSet_i^R$}.\\
    0, & \text{otherwise}.
  \end{cases} $$  
The second constraint, Constraint~\eqref{BP3}, forces the solver to choose at least one column for each client. For the linear relaxation, the sum of ''probabilities'' of using columns for a particular client $\client_{i}$ from $\columnSet^R_i$ must be greater than or equal to 1. Note that this sum is always 1 in the optimal solution of the initial problem, as criterion minimization pushes it down. This trick reduces computation time, since a feasible solution can be found earlier in the branching tree of the master model, resulting in a better upper bound.
Having a good upper bound earlier allows bounding more efficiently, which reduces computation time.

\begin{equation} \label{BP3}
\sum_{\pattern_{\patternIndex} \in \columnSet_i^R} \belongs_{i,\patternIndex} \cdot \assignment_{i,\patternIndex} \geq 1, \qquad \client_{i} \in \clientSet.
\end{equation}



\subsection{Sub-Model Formulation}
\label{BP_SM} 

The main purpose of the sub-model is to generate the columns $\columnSet^R$ for the master model. Apart from that, the sub-model is used to check optimality of the restricted master model $MM(\columnSet^R)$. The following condition guarantees optimality of the solution obtained by $MM(\columnSet^R)$ (see~\cite{Vanderbei96} for the proof) and it is used for constructing the sub-model formulation:

\vspace{\beforeSpace}
\begin{condition}[Optimality condition]
\label{cond:optimality}
A solution to a linear relaxation of $MM(\columnSet^R)$ is optimal if and only if there is no column $\pattern_{\patternIndex} \in \columnSet$ that is infeasible for the dual model to the linear relaxation of $MM(\columnSet^R)$. 
\end{condition}
\vspace{\afterSpace}

The notion of duality between two LP models is described in~\cite{Feilet10}. Basically, it exchanges constraints and variables in their LP formulations. Note that the procedure of constructing a \textit{dual master model} formulation $DMM(\columnSet^R)$ to $MM(\columnSet^R)$ does not contain any design decisions and follows automatically from the formulation of $MM(\columnSet^R)$. 

The formulation of $DMM(\columnSet^R)$ is given below, where $\dualsOne_j$ are dual variables that correspond to the set of Constraints~\eqref{BP2} and $\dualsTwo_i$ are dual variables for Constraints~\eqref{BP3} of the master model. $DMM(\columnSet^R)$ aims to maximize Criterion~\eqref{BP4} with respect to Constraints~\eqref{BP5}--\eqref{BP7}.

\begin{equation}\label{BP4}
Maximize: -\sum_{j \in \frameSizeSet}\dualsOne_j + \sum_{\client_{i} \in \clientSet } \dualsTwo_i.
\end{equation} 
\emph{subject to}:

\begin{equation} \label{BP5}
-\sum_{j \in \frameSizeSet} \scheduleBP_{i,\patternIndex}^j \cdot \dualsOne_j  + \belongs_{i,\patternIndex} \cdot \dualsTwo_i \leq \frac{\slots_{i,\patternIndex}}{ \frameSize},  \qquad \client_{i} \in \clientSet,\\ \pattern_{\patternIndex} \in \columnSet_i,
\end{equation}
\begin{equation} \label{BP6}
\dualsOne_j  \leq M', \qquad j \in \frameSizeSet
\end{equation}
\begin{equation} \label{BP7}
\dualsOne_j  \geq 0, \qquad j \in \frameSizeSet
\end{equation}

Variables $\dualsOne_j$ and $\dualsTwo_i$ are called \textit{shadow prices}. In terms of the TCP/LR-F problem, they can be interpreted as how much the value of Criterion~\eqref{BP1} of the master model would decrease if the corresponding constraints of $MM$ are relaxed. Particularly, $\dualsOne_j$ indicates a potential gain in terms of criterion value if slot $j$ is allowed to be allocated twice (in general, allowing slot $j$ to be allocated $k$ times, the criterion reduces by $(k - 1) \cdot \dualsOne_j$). Meanwhile, $\dualsTwo_i$ is the price of having one column for client $\client_{i}$ both in terms of its allocated rate $\bandwidth_i$ and its slot assignment, i.e. the criterion value~\eqref{BP1} would reduce by $\dualsTwo_i$ if client $\client_{i}$ could have no columns in the final solution.

As already stated, the sub-model generates new promising columns for an individual client. Promising are those that violate feasibility of $DMM(\columnSet^R)$ according to the optimality stated in Condition~\ref{cond:optimality}. The negated value by which Constraints~\eqref{BP5}--\eqref{BP7} are violated by a new column $\pattern_{\patternIndex}$ is called the \textit{reduced cost}. In TCP/LR-F, the reduced cost value can be interpreted as how much the cost of having certain column in a final solution (in terms of the criterion value) must be reduced before it is included in the optimal solution. From the point of view of Condition~\ref{cond:optimality}, negative reduced cost means the current solution of $MM(\columnSet^R)$ is not optimal. Basically, the sub-model searches for a new column, defined by Equations~\eqref{con:tmind2},~\eqref{con:tmind3} and~\eqref{con:tmind4}. To find such a column with minimal reduced cost, it uses the ILP model from Section~\ref{model} for a single client. This is how the ILP model is used as a building block in a bigger framework. 

Having minimal reduced cost does not necessarily mean that the column brings the result towards an optimal solution. However, choosing the column with the minimal reduced cost is a heuristic that behaves very well in practice~\cite{Feilet10} and eventually leads to the optimal solution due to~Condition~\ref{cond:optimality}.

Next, the formulation of the sub-model is given. Since it is required to minimize the reduced cost, violation of Constraints~\eqref{BP5}-\eqref{BP7} needs to be formulated in terms of the sub-model variables. Remember that the values $\dualsOne_j$ and $\dualsTwo_i$ are constants from the sub-model point of view, since they are obtained from the master model after Step 2 in Figure~\ref{fig:BPScheme}. Thus, Constraints~\eqref{BP6} and~\eqref{BP7} are satisfied by default and it is sufficient to consider violation of Constraint~\eqref{BP5} only. 

Looking closely at Constraint~\eqref{BP5}, it is clear that the sub-model only needs to determine slot allocation for the given client $\client_{i}$, i.e. variables $\scheduleBP_{i,\patternIndex}^j$ are in fact the same as $\scheduleMod_{i}^{j}$ from the ILP model, previously described in Section~\ref{model}. Furthermore, $\slots_{i,\patternIndex}$ is the number of allocated slots in column $\patternIndex$, which implies $\slots_{i,\patternIndex} = \slots_{i} =\sum_{j \in \frameSizeSet}\scheduleMod_{i}^{j}$. Finally, $\belongs_{i,\patternIndex}$ is always 1 in the sub-model, since the schedule is constructed for client $\client_{i}$. Thus, the sub-model for client $\client_{i}$ has the criterion $\critSM$~\eqref{BP8_1}, which is the reduced price expression.

\begin{equation}\label{BP8_1}
Minimize: \sum_{j \in \frameSizeSet} \scheduleMod_{i}^{j} \cdot \dualsOne_j + \frac{\sum_{j \in \frameSizeSet}  \scheduleMod_{i}^{j}}{\frameSize} - \dualsTwo_i = \critSM.
\end{equation} 


The constraints of the sub-model duplicate the constraints of the ILP model in Section~\ref{model} for $C = \{\client_{i}\}$, i.e. considering client $\client_{i}$ only. Constraint~\eqref{BP9} states that the bandwidth requirement must be fulfilled, Constraint~\eqref{BP10} computes the points of the worst-case provided service line and Constraint~\eqref{BP11} guarantees satisfying the service latency requirement for the given client according to Definition~\ref{def:lr_server}. Moreover, all the optimizations for the ILP model, described in Section~\ref{model}, are used for the sub-model as well. 

 \begin{equation} \label{BP9}
\sum_{j = 1}^{\frameSize}  \scheduleMod_{i}^{j} \geq \frameSize \cdot \requiredBandwidth_i
\end{equation}
 \begin{equation} \label{BP10}
\minNumberDoneWork_{i}^{j} \leq \sum_{l = k}^{(k - j) \bmod \frameSize}  \scheduleMod_{i}^{j}, \: k \in \frameSizeSet, \: j \in \frameSizeSet
\end{equation}
 \begin{equation} \label{BP11}
\minNumberDoneWork_{i}^{j} \geq \requiredBandwidth_i\cdot  (j - \requiredLatency_i), \; j \in \frameSizeSet.
\end{equation}

Thus, here the ILP model, formulated in Section~\ref{model}, is used as a piece in a larger framework to solve large problems more efficiently due to the ILP decomposition, i.e. instead of solving a large model that considers all clients at the same time, the master model and the sub-model solve smaller sub-problems.
A small illustrative example of the column generation algorithm on a problem instance with 2 clients can be found in the appendix. 

\subsection{Branching Strategy}
\label{BP_branching}

The third and last main component of the branch-and-price approach is the branch-and-bound procedure. Here, we focus on the \emph{branching strategy} used. Branch-and-price approaches usually use the 0/1 branching scheme, i.e. some binary variable is set either to 0 or to 1 in two child nodes. First of all, the important decision is how to choose variables to branch on. Branching on master model variables ($\assignment_{i, p}$) is not effective since new columns are added in each iteration and the number of decisions to make increases with added columns, which results in a large decision tree and long computation time.
Moreover, it is more efficient to consider variables of the sub-model as the branching variables, since it influences several columns at once.

Generally, branching adopts a depth-first search. Note that two branching decisions must be made: which \textit{client} to allocate to which \textit{slot}. Two branching strategies show good results for different sizes of problems. Both strategies start by sorting clients in ascending order according to their service latency requirements to begin branching from the clients with more critical requirements. 

The first branching strategy has experimentally shown good results for smaller problems (up to 16-32 clients). It fixes $\scheduleMod_{i}^{j}$ in the following order, assuming the above described sorting of the clients: $ \scheduleMod_{1}^{2}, \cdots, \scheduleMod_{1}^{\frameSize}, \scheduleMod_{2}^{1},\cdots,\scheduleMod_{\numClients}^{\frameSize}$, i.e. slots are assigned sequentially from left to right for each client. Note that the first slot is always allocated to the client with the tightest service latency requirements to reduce symmetry in the solution space (similar to the ILP from Section~\ref{model}). This branching first makes decisions of type $ \scheduleMod_{i}^{j} = 0$, i.e. branching goes first to the branch where slot $j$ is forbidden to be allocated to client $\client_{i}$ and only then to the branch where it is fixed to be allocated. The reason for this is to obtain a feasible solution as fast as possible, since negative decisions are less likely to eliminate any column in $\columnSet^R$ and therefore do not require time to be spent running column generation. Moreover, going through the solution space in a systematic manner reduces the number of explored symmetrical solutions.

The second branching strategy shows better results for larger use-cases (more than 16-32 clients). It begins by computing the total ''probabilities'' of including each slot $j \in \frameSizeSet$ in the solution for all its columns,  $\assignment_{j}^{\text{total}}$. For example, for the first client this means computing $\assignment_{j}^{\text{total}} = \sum_{\pattern_{\patternIndex} \in \columnSet^R_1: \scheduleBP_{1,\patternIndex}^j = 1} \assignment_{1, \patternIndex}$. Then it chooses the non-decided slot $j_c$ with maximum total ''probability'' of being in the solution, $j_c = argmax_{j \in \frameSizeSet}  \: \assignment_{j}^{\text{total}}$ and branches on it. In case there are multiple slots with the same maximum $\assignment_{j}^{\text{total}}$, the leftmost slot $j_c$ is chosen. While branching on variable $\scheduleMod_{j_c}^{1}$, it goes to the branch $\scheduleMod_{j_c}^{1} = 1$ first. If the first client is fully decided or all the ''probabilities'' of undecided slots are zero, the procedure continues with the following client. This branching strategy typically leads to a feasible solution early on as slots with non-zero probability are the promising ones to allocate.

\begin{figure*}
\centering
\includegraphics[width=2\columnwidth]{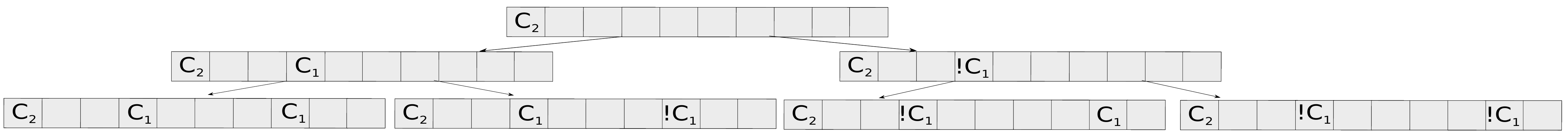}
\caption{First three layers of the example branching tree.}
\label{fig:BranchingScheme}
\end{figure*} 

An example with the first three layers of a branching tree using the second branching strategy is shown in Figure~\ref{fig:BranchingScheme}. It assumes the root node contains the set of columns $ \columnSet^R$ from Figure~\ref{fig:column} and the master model resulted in the variables $\assignment_{1,1} = 0.2$, $\assignment_{1,2} = 0.8$, $\assignment_{2,1} = 0.7$, $\assignment_{2,2} = 0.3$ in Step~2 of Figure~\ref{fig:BPScheme}. For the first client there are 4 slots $j = (4, 8, 9, 10)$ with the same $\assignment_{j}^{\text{total}} = 1$ and neither of them is decided yet. Therefore, the branching decision in the root node is to use client $\client_{i}= 1$ and slot $j = 4$ for branching. As a consequence, the left child has a partial solution, where $\scheduleMod_{1}^{4} = 1$ and the right child has  $\scheduleMod_{1}^{4} = 0$ ($!\client_{1}$). Note that column generation in the nodes of the second layer starts with a changed set of initial columns. Therefore, the branching decisions are made based on different values of $\assignment_{i,j}$. 
 
\section{Computation Time Optimizations}
\label{BP_details}

The main components of the branch-and-price approach are specified for TCP/LR-F in the previous section. However, there are plenty of opportunities for reducing the computation time of the approach. Numerous experiments and observations were done and this section presents the five most successful optimizations to reduce computation time. 

First of all, starting the branch-and-price approach from a feasible initial solution obtained by a \emph{heuristic}
implies a significant reduction of the computation time, since having a good upper bound on the solution in the beginning reduces the size of the branching tree. 
If the heuristic is good, it may often be able to find the optimal solution by itself. In this case, the branch-and-price approach only has to prove
the optimality in the root node of the branching tree, considerably reducing the computation time. This is the case for our heuristic, later presented
in Section~\ref{heuristic}.

Secondly, since the sub-model for non-latency-dominated clients requires $\frameSize \, ^2$ constraints for checking service latency guarantees, \emph{lazy constraints}~\cite{cplex} are used. The basic idea is to initially formulate the problem only with the most esssential constraints, omitting those that are only rarely violated. These other constraints are checked and added one-by-one to the model only if the solution violates any of them. This process is done in an efficient way. Instead of always starting from the beginning of the branching tree, it continues search from the place in the tree it last finished. We apply this trick to the sub-model, which we initially formulate assuming all clients are latency dominated, i.e. having only $\frameSize$ Constraints~\eqref{con:tmind3} and~\eqref{con:tmind4} for $j = \floor{\requiredLatency_i} + 1$. Next, for bandwidth-dominated clients, the solution is checked for satisfaction of Constraints~\eqref{con:tmind3} and~\eqref{con:tmind4} for $j > \floor{\requiredLatency_i} + 1$. If a constraint is violated, it is added to the model and the solver continues its search until it is finished. The key idea behind this optimization is to exploit that although service latency is not always the largest gap between two consecutively allocated slots to the same client, it often is, and the extra constraints
for bandwidth-dominated clients are hence typically not necessary. 


The next optimization also concerns the branch-and-bound part. We use problem-specific information to \emph{set the bounding condition} more effectively. The bounding condition in every node is set to be $\LBOnCriterion >  \UBOnCriterion- 1/\frameSize$, where the discretization step $1/\frameSize$ is subtracted. The reason is that unless there is at least one slot less, which is exactly  $1/\frameSize$ in terms of utilization, it is not an improved solution to the problem. 

Column generation often suffers from a so called "tailing-off" effect, i.e. at the moment the reduced prices of sub-models are close to zero it starts to take a lot of iterations to converge to exact zeros. To deal with this issue, the fourth optimization, \emph{Lagrangian relaxation}, is introduced. It estimates the lower bound  $\LBOnCriterion$ on $\Criterium$ inside the column generation loop by Equation~\eqref{BP13}, before the master model is solved to optimum. Thus, using Lagrangian relaxation stops the column generation process before Condition~\ref{cond:optimality} holds and closes the node, saving computation time. It is applicable when either the estimated lower bound $\EstimatedLBOnCriterion \geq \UBOnCriterion$
or if the estimated lower bound $\EstimatedLBOnCriterion$ and the one given by the master model after previous iteration $\critMM_{curr}$ discretize to the same number of allocated slots. Equation~\eqref{BP13} computes the lower bound on the criterion value of the master model after each iteration inside the column generation procedure after Step~3 of Figure~\ref{fig:BPScheme}. The estimation is the current criterion value of the master model $\critMM_{curr} $ plus the sum of all of the reduced prices $\critSM_i$ of the sub-models~\cite{Huisman03}. The Lagrangian relaxation can never estimate $\LBOnCriterion$ incorrectly~\cite{Huisman03}, i.e. it does not break optimality.

\begin{equation} \label{BP13}
\EstimatedLBOnCriterion = \critMM_{curr} + \sum_{\client_{i} \in \clientSet} \critSM_i
\end{equation}

Note that each time the Lagrangian relaxation is applied, all the sub-models must be solved to optimum. Hence, running Lagrangian relaxation each iteration of column generation could increase the total computation time, so it is an important decision when to start this process. 
In our approach, Lagrangian relaxation is performed each $\numClients$ iterations, where $\numClients$ is the number of clients. This design decision is motivated by that when all $\numClients$ clients have run, there is higher chance that the reduced prices have changed significantly and there is space for the Lagrangian relaxation to close the node.

The fifth and the final optimization of the computation time is \emph{solution completion} by the ILP from Section~\ref{model}. Sometimes branching goes deep enough so that it is possible to reduce computation time by launching the ILP model to find a schedule for all clients simultaneously. It is done when some percent of positive or negative decisions has been made. Our experiments have shown that for different sizes of problems these parameters should be set differently. For example, it is necessary to increase them with increasing number of clients, since larger problems could require running the ILP for a long time. Thus, nodes at this depth are solved to optimality using the ILP model and are closed afterwards. This procedure is done by taking the decisions that have already been made, fixing corresponding variables in the ILP model and solving this problem with an ILP solver.

\section{Heuristic Approach}
\label{heuristic}

Although the proposed exact approaches solve TCP/LR-F optimally, it is sometimes acceptable to sacrifice the quality of the solution in order to reduce computation time. Moreover, as mentioned earlier, the branch-and-price approach can use a heuristic solution to compute good initial columns that reduce the total computation time. Thus, the purpose of this section is to present a heuristic that solves the TCP/LR-F problem.

Heuristics of the constructive type (ones that construct a solution step-by-step) for the considered TCP/LR-F problem lack a good strategy to backtrack from low quality solutions and this is the reason we propose a \emph{generative heuristic} that generates a complete solution at once. Although the generated solution may initially be infeasible, the heuristic gradually changes it towards a feasible one.

The proposed heuristic exploits the sub-model, previously described in Section~\ref{BP_SM}. It is used in combination with the lazy constraints presented in Section~\ref{BP_details} and with the optimality gap set to 5\%, i.e. it is not necessarily the optimal solution that is returned by the solver, but one that is no further than 5\% relative distance from the result of the linear relaxation of the problem. These improvements significantly reduce the computation time. 

The heuristic constructs the schedule by iteratively running the sub-model for different clients in a cyclic manner. Remember that the sub-model aims to minimize $\critSM = \sum_{j \in \frameSizeSet} \scheduleMod_{i}^{j} \cdot \dualsOne_j + \frac{\sum_{j \in \frameSizeSet}  \scheduleMod_{i}^{j}}{\frameSize} - \dualsTwo_i$, where the dual price coefficients $\dualsOne_j$ control allocation of slot $j$. In this heuristic, $\dualsOne_j$ is not coming from the master model, but is determined by the solutions from previous iterations. Note that in the column generation procedure, this is done by the master model in Step 2 of Figure~\ref{fig:BPScheme}. Here, the heuristic substitutes Steps 2 and 7 of Figure~\ref{fig:BPScheme} with a procedure that assigns appropriate coefficients $\dualsOne_j$. Meanwhile the dual price $\dualsTwo_i$ is omitted in the heuristic as it is a constant in a sub-model and constant plays no role in minimization.

Algorithm~\ref{alg:Heuristic} shows the proposed heuristic. The number of clients $\numClients$, service latency $\requiredLatency$ and bandwidth requirements $\requiredBandwidth$ are used as input. Furthermore, there are two parameters of the heuristic, a coefficient $\alpha$, which controls the speed of convergence to the final solution, and the maximum number of iterations of the sub-model $N_{iter}^{max}$. Each iteration includes two main steps: first coefficients $\dualsOne_j$ are computed on Line~4 (explained later) and then the sub-model for client $\client_{i}$ is launched on Line~5. Note that the current solution is the schedule constructed out of the $\numClients$ last created columns for individual clients. The heuristic stops either when the maximum number of iterations, $N_{iter}^{max}$, is reached or if the current solution $ \scheduleMod_{i}^{j, curr}$ is collision-free, i.e. there are no two clients that share a slot in the current solution. 

\begin{algorithm}[htb]
\caption{The proposed generative heuristic}
\label{alg:Heuristic}
\begin{algorithmic}[1]
\STATE Inputs: $\numClients, \requiredLatency, \requiredBandwidth, \alpha, N_{iter}^{max}$
\STATE $N_{iter} = 0$, $i = 1$, $\scheduleMod_{i}^{j, curr} = 0 \; \forall \client_{i} \in \clientSet, \: j \in \frameSizeSet.$ 
\WHILE {$N_{iter} <  N_{iter}^{max}$ and $\scheduleMod_{i}^{ j, curr}$ has collisions}
\STATE $\dualsOne$ = ComputeCoefficients($\client_{i}$, $\alpha$)
\STATE $\scheduleMod_{i}^{j, curr}$ = SubModel($\requiredLatency_i, \requiredBandwidth_i$, $\dualsOne$)
\STATE $N_{iter} = N_{iter} + 1$
\STATE $i = (i \mod \numClients) + 1$ 
\ENDWHILE
\STATE Output: $\scheduleMod_{i}^{j, curr}$
\end{algorithmic}
\end{algorithm}

The core of the heuristic is fast assignment of coefficients $\dualsOne_j$ to each slot for a given client, such that if multiple clients allocate the same slot, some of them will change their allocation. As we are minimizing the criterion value, a higher value of the coefficient means it is less desirable for a client to allocate slot $j$ ($\dualsOne_j \in [0.9, 2.5], j \in \frameSizeSet$). The procedure of assigning coefficients $\dualsOne_j$ for client $\client_{i}$ is presented in Algorithm~\ref{alg:ComputeCoefficients}. The algorithm considers four mutually exclusive and jointly exhaustive cases that are detailed below:

\begin{enumerate}
\item \emph{Slot $j$ is allocated to some client $\client_{k} \neq \client_{i}$ and not allocated to client $\client_{i}$} in the current schedule $\scheduleMod_{i}^{j, curr}$. Generally, in this situation slot $j$ should not be allocated to client $\client_{i}$ to avoid conflict, but in early stages of the heuristic this slot can be used if necessary, since it is not known in advance which allocation is better. The corresponding coefficient is assigned $\dualsOne_j = min(2,\ 1 + d_{j,i} \cdot \alpha) $, where $d_{j,i}$ is the number of times slot $j$ was allocated in the previous iterations to any client $\client_{k}$, $k \neq i$. The coefficient $d_{j,i}$ hence adds state to the iterative algorithm: the more times the slot was allocated to other clients before, the less attractive it is to allocate this slot to client $\client_{i}$. Furthermore, the higher the value of coefficient $\alpha$, the faster the schedule converges, since $d_{j,i} \cdot \alpha$ becomes larger. The upper bound of 2 limits the maximal penalization. 

\item \emph{Slot $j$ is allocated to client $\client_{i}$ in $\scheduleMod_{i}^{j, curr}$ and there is no conflict}. The constant value of $\dualsOne_i = 0.9$ is chosen here, since the algorithm prefers not to change the clients allocation unless necessary. In case this value would be set too low, the algorithm will tend to allocate a new slot rather than changing the allocation of the currently assigned slots.

\item \emph{Slot $j$ is allocated in $\scheduleMod_{i}^{j, curr}$ to client $\client_{i}$ and there is a conflict}. All but one of the conflicting clients should leave the slot. However, it is not clear in advance which client should have the slot. Therefore randomness is introduced here - the coefficient in this case is selected from uniformly distributed numbers between 1 and 2.5.

\item In case \emph{$j$ is not allocated} in the current schedule, a value $\dualsOne_j = 1.0$ is assigned. 
\end{enumerate}

\begin{algorithm}[htb]
\caption{ComputeCoefficients}
\label{alg:ComputeCoefficients}
\begin{algorithmic}[1]
\STATE Inputs: $\client_{i}; \alpha;$
\FORALL {$j \in \frameSizeSet$} 
   \STATE {$d_{j,i}$ = number of times slot j was allocated in the previous iterations to any client $\client_{k} \neq \client_{i}$}
   \STATE {$\dualsOne_j = \begin{cases}
    min(2,1 + d_{j,i} \cdot \alpha), & \text{if $\scheduleMod_{i}^{j,curr} = 0$, $\scheduleMod_{k}^{j,curr} = 1$,}\\
    									& \ \ \ \ \ \ \ \  \text{for some $k \neq i$. (1)}\\
    0.9, & \text{if $\scheduleMod_{i}^{j,curr} = 1$, $\scheduleMod_{k}^{j, curr} = 0$}, \\
									&\ \ \ \ \ \ \ \  \text{for every $ k \neq i$. (2)}\\
    1 + rand() \cdot 1.5, & \text{if $\scheduleMod_{i}^{j,curr} = 1$, $\scheduleMod_{k}^{j, curr} = 1$,}\\
 									& \ \ \ \ \ \ \ \   \text{for some $k \neq i$. (3)}\\
     1, & \text{if $\scheduleMod_{k}^{j,curr} = 0$}\\
     									&   \  \text{for every $k = 1,...,\numClients$. (4)}
  \end{cases} $}
\ENDFOR
\STATE Output: $\dualsOne$
\end{algorithmic}
\end{algorithm}


\section{Experimental Results}
\label{experiments}
This section experimentally evaluates and compares the TDM configuration methodology based on the branch-and-price approach and the proposed heuristic. First, the experimental setup is explained, followed by an experiment that compares the proposed branch-and-price and heuristic to the existing ILP from Section~\ref{model} in terms of scalability. Furthermore, it shows the trade-off between criterion value and computation time for the heuristic approach.

\subsection{Experimental Setup}

\begin{table*}[h]
\centering
\rowcolors{1}{}{gray!35}
\caption{Parameters for use-case generation}
\label{table:instance_generation}
\begin{tabular}{|c|c|c|c|c|c|c|}
\hline
\multirow{2}{*}{\emph{Clients}} & \multicolumn{2}{c|}{\emph{Bandwidth-dominated}} & \multicolumn{2}{c|}{\emph{Latency-dominated}} &  \multicolumn{2}{c|}{\emph{Mixed-dominated}} \TBstrut\\ \cline{2-7} 
    & $\rateParam$             & $\latencyParam$        & $\rateParam$                 & $\latencyParam$         & $\rateParam$                 & $\latencyParam$      \TBstrut\\ \hline
8   & [0.06, 0.16]  & [0.6, 0.9]  & [0.02, 0.07]  & [1.6, 3.3]  & [0.06, 0.14] & [0.95, 1.4] \TBstrut\\ \hline
16  & [0.03, 0.08]  & [0.5, 0.75] & [0.01, 0.035]  & [1.58, 3.26]  & [0.03, 0.07] & [0.9, 1.3]  \TBstrut\\ \hline
32  & [0.015, 0.04] & [0.4, 0.6]  & [0.005, 0.0175] & [1.56, 3.22]  & [0.015, 0.035] & [0.85, 1.2]  \TBstrut\\ \hline
64  & [0.0075, 0.02] & [0.3, 0.45]  & [0.0025, 0.00875] & [1.54, 3.18]  & [0.0075, 0.0175] & [0.8, 1.1]  \TBstrut\\ \hline
128  & [0.00375, 0.01] & [0.2, 0.3]  & [0.00125, 0.004375] & [1.52, 3.14]  & [0.00375, 0.00875] & [0.75, 1.0]  \TBstrut\\ \hline
\end{tabular}
\end{table*}

Experiments are performed using three sets of $5 \times 200$ synthetic use-cases, each comprising 8, 16, 32 and 64 or 128 real-time clients on one resource. The three sets are bandwidth-dominated, latency-dominated and mixed-dominated use-cases. The concepts of latency-dominated and bandwidth-dominated clients were previously introduced in Section~\ref{model}. A mixed-dominated client is one that requires approximately equal allocated rate to satisfy both service latency and bandwidth requirements according to the right or left part of Equation~\eqref{con:tmind7}, respectively. A mixed-dominated use-case comprises only mixed-dominated clients. This type of use-cases was not considered in~\cite{Akesson15} due to high time complexity. The reason for looking at this group of instances is that the problem is more difficult than in bandwidth-dominated use-cases, but unlike latency-dominated cases, constraints cannot be removed by the computation time optimizations in Section~\ref{model}.
The reason for evaluating these three classes is to show the impact of the requirements on the computation time of the proposed branch-and-price approach, as well as fairly evaluate the efficiency of the heuristic. We proceed by explaining how bandwidth and service latency requirements are generated for the three sets.

Parameters for synthetic use-case generation are given in Table~\ref{table:instance_generation}. Firstly, bandwidth requirements of each client in a use-case are generated. Here, $\rateParam$ is an interval from which bandwidth requirements for each client are uniformly drawn. The use-case is accepted if the total required rate of all clients is in the range [0.8, 0.95] for bandwidth-dominated use-cases, [0.35, 0.5] for latency-dominated use-cases and [0.7, 0.9] for mixed-dominated use-cases. Otherwise, it is discarded and the generation process restarts. 
The interval of acceptance is lower for both mixed-dominated and latency-dominated sets to leave space for over-allocation to satisfy the tighter service latency requirements. Each time the number of clients is doubled, the range of bandwidth requirements is divided by 2. This is to make sure the total load is comparable across use-cases with different number of clients, which is required to fairly evaluate scalability.

Service latency requirements are uniformly distributed according to $\frac{1}{\latencyParam \cdot \requiredBandwidth}$, where a larger value of $\latencyParam$ indicates a tighter requirement.  The $\latencyParam$ values are given in Table~\ref{table:instance_generation}.
The reduction of service latency requirements with increasing number of clients is empirically determined to provide instances with  comparable difficulty by having similar total allocated rates for the final optimal schedules. Lastly, if the total possible load due to the service latency requirements (the right part of Equation~\eqref{con:tmind7}) is outside the interval $[0.75, 0.95]$ and $[0.7, 0.9]$, for latency-dominated and mixed-dominated use-cases, respectively, new latency requirements for the use-case are generated. For all the sets, generated use-cases that are found infeasible using the optimal approach are discarded and replaced to ensure a sufficient number of feasible use-cases. Finally, the frame size is set to $\frameSize = \numClients \cdot 8$ to make sure that the number of slots available to each client is constant across the experiment. 

All in all, this generation process ensures that all use-cases are feasible, have comparable difficulty, and that all clients in the three sets have the desirable dominating property. Experiments were executed on a high-performance server equipped with 2x Intel Xeon X5570 (2.93~GHz, 20~cores total) and 100~GB memory. The ILP model and ILP part of branch-and-price and the heuristic were implemented in IBM ILOG CPLEX Optimization Studio~12.5.1 and solved with the CPLEX solver using concert technology for the latter two. The branch-and-price and heuristic approaches were implemented in JAVA.

\subsection{Results}

The experiments evaluate the scalability of the proposed branch-and-price approach and the trade-off between computation time and the total rate (the criterion) allocated to 8, 16, 32, 64 and 128 real-time clients for the optimal and heuristic approaches. Moreover, it compares the proposed approaches with already existing exact (the ILP formulation from Section~\ref{model}) and heuristic (continuous allocation) strategies. A time limit of 3 000 seconds per use-case was set in order to obtain the results of the experiments in reasonable time. Furthermore, to get higher quality  solutions at expense of increased computation time, the heuristic was launched 8 times for latency-dominated and mixed-dominated use-cases to exploit the random component of the heuristic.
Note that the heuristic is only run once for bandwidth-dominated use-cases, since these are easier for the heuristic to solve. The heuristic parameters were set to $\alpha = 0.1$, $N_{iter}^{max} = 250$. Besides, the used branching strategy for the use-cases with 8 and 16 clients is the first (consecutive) one, while for the use-cases with 32, 64 and 128 clients the second (maximum total probability) one was selected. Furthermore, after 10, 30, 60, 80, and 95\% of positively decided slot allocations and 40, 100, 120, 260 and 300\% of negative decisions in terms of time slots (with 100\% being $\frameSize$) is done for the use-cases with 8, 16, 32, 64 and 128 clients, respectively, the completion by the ILP model is launched. These numbers were empirically determined to provide a reasonable trade-off between computation time and quality of the solution. 

\vspace{3mm}

\subsubsection{Bandwidth-dominated use-cases}

\begin{figure}[h]
\centering
\epsfig{file=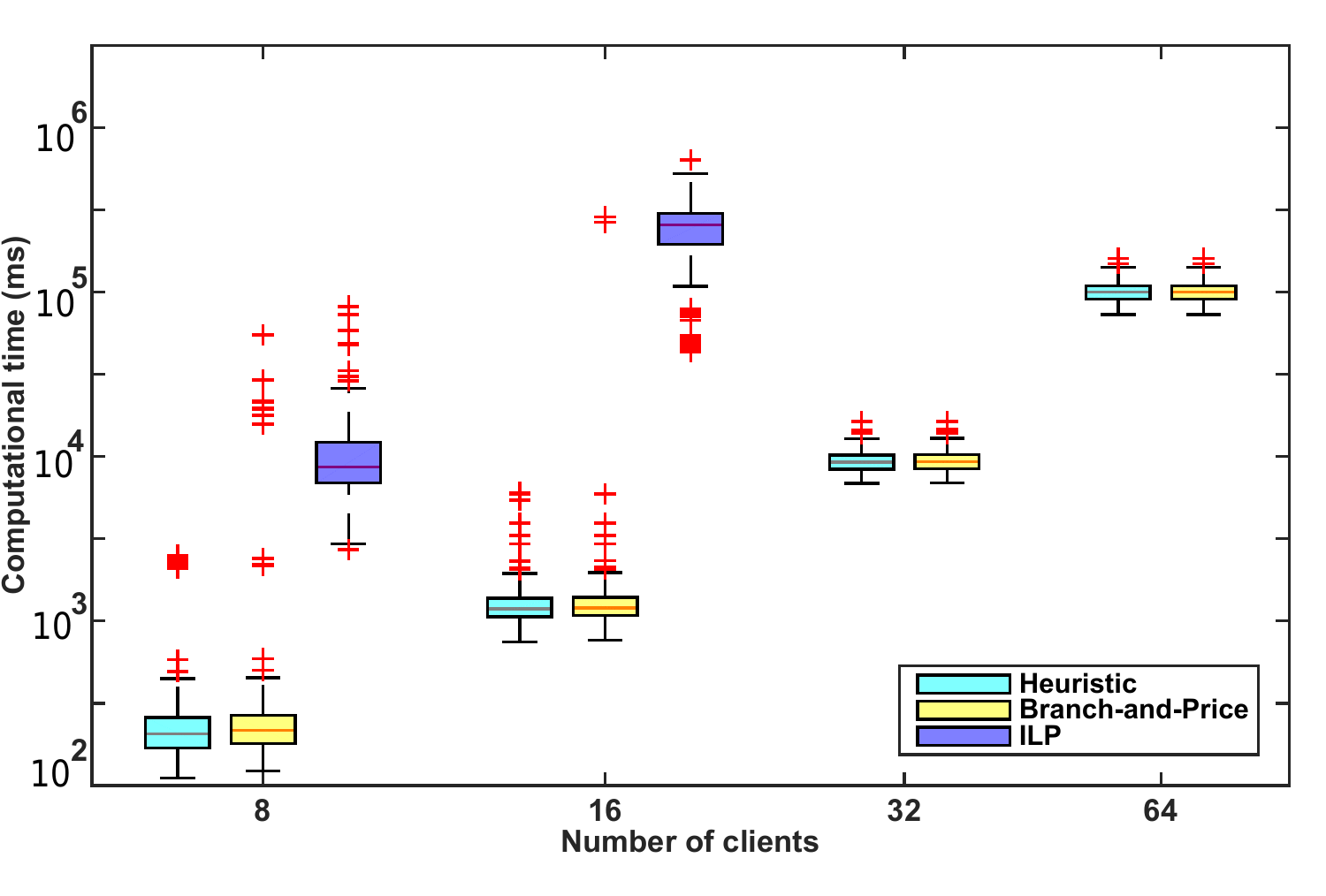,width=1\columnwidth}
\caption{Computation time distribution for the bandwidth-dominated use-cases.}
\label{fig:timeBandDominated}
\end{figure}

Figure~\ref{fig:timeBandDominated} shows the vertical axis in logarithmic scale of the computation time for the bandwidth-dominated use-cases for the heuristic, branch-and-price and ILP approaches for 8, 16, 32, 64 and 128 clients, respectively. The ILP struggles to scale to use-cases with 32 clients, as 52 out of 200 use-cases are not solved to optimality within the given time limit, resulting in a failure rate of above 25\%. Therefore, the results for 64 and 128 clients are only represented by the heuristic on the left and branch-and-price on the right. The results show that the branch-and-price approach significantly outperforms the ILP model, scaling well to the use-cases with 32, 64 and 128 clients. More specifically, branch-and-price require 10\% (4 minutes) and 2\% (14 minutes) of the ILP computation time for 8 and 16 clients, respectively. Moreover, it solves the use-cases with 32, 64 and 128 clients in 0.5, 6 and 29 hours, respectively, resulting in less than 9 minutes computation time per use-case on average for the use-cases with 128 clients. The quality of the solution is shown in Table~\ref{table:failure}, where the first number is the number of failures and the second one is the average distance (excluding the failures) from the best obtained solution. Failure is defined as no feasible solution for the heuristic and no optimal solution for the exact approaches within the time limit. The results for 128 clients are not presented in the table, since they are identical to the results for 64 clients. The reason for this is that for both types of use-cases, the heuristic was able to solve all 200 use-cases to optimality and branch-and-price only had to prove optimality of the solutions.


\begin{table*}[h]
\centering
\rowcolors{1}{}{gray!35}
\caption{Number of failures and average distance from the best obtained solution. (BD -- bandwidth-dominated, LD -- latency-dominated and MD -- mixed-dominated use-cases)}
\label{table:failure}
\begin{threeparttable}
\begin{tabular}{|c|c|c|c|c|c|c|c|c|c|c|c|c|}
\hline
\multirow{2}{*}{\emph{Clients}} & \multicolumn{3}{c|}{\emph{8 clients}} & \multicolumn{3}{c|}{\emph{16 clients }} &  \multicolumn{3}{c|}{\emph{32 clients}} &   \multicolumn{3}{c|}{\emph{64 clients}}  \TBstrut\\ \cline{2-13} 
    & \emph{BD}             & \emph{LD}        & \emph{MD}    & \emph{BD}             & \emph{LD}        & \emph{MD} & \emph{BD}             & \emph{LD}        & \emph{MD}  & \emph{BD}             & \emph{LD}        & \emph{MD}      \TBstrut\\ \hline
    
ILP   &  0/0  & 0/0 & 1/0 & 0/0 & 1/0 & 1/0 & 52/0.0005 & 0/0 & 172/0 & - & 0/0 & - \TBstrut\\ \hline
Branch-and-Price  &  0/0 & 6/0 & 3/0 & 0/0 & 1/0 & 0/0 & 0/0 & 0/0 & 0/0 & 0/0 & 5/0 &0/0 \TBstrut\\ \hline
Heuristic  &  8/0 & 46/0.01 & 23/0.01 & 2/0 & 76/0.006 & 26/0.003 & 0/0 & 71/0.001 & 8/0.0001 & 0/0 & 61/0.0001 & 0/0 \TBstrut\\ \hline
\end{tabular}


\end{threeparttable}
\end{table*}

Comparing the heuristic and the branch-and-price approach for the bandwidth-dominated use-cases, the results indicate a slight time reduction with some loss in the quality of the solution for the heuristic. The distributions of the computation time of both the heuristic and the branch-and-price approach for 8, 16, 32, 64 and 128 clients look similar with exception of the use-cases marked as plus signs, for which the heuristic was not able to find any feasible solutions. However, the results in Table~\ref{table:failure} show that when the heuristic managed to find feasible solutions, it always found the optimal one. 
The second issue that seems suspicious is the visible similarity of the distribution of computation time of both the heuristic and branch-and-price approaches, which is caused by using $log_{10}$ scale, where the small difference becomes invisible. In reality, the total computation time of the heuristic and the branch-and-price differ. The heuristic runs in 25\%, 29\%, 96\%, 99\% and 99\% of the branch-and-price computation time for 8, 16, 32, 64 and 128 clients, respectively. Such a similarity for 32, 64 and 128 clients is a result of the heuristic being successful in all use-cases, which could be caused by having more space for allocation without collisions when the frame size is longer.

To push the limits of the branch-and-price approach and find the maximum number of clients it can manage, a use-case with 256 clients was generated by following the rules in Table~\ref{table:instance_generation}. However, because of the memory limit of 100~GB, branch-and-price was not able to finish the run for a single use-case. The current limits for our approach
is hence somewhere between 128 and 256 bandwidth-dominated clients in a use-case.

\emph{This experiment shows that for bandwidth-dominated use-cases the branch-and-price approach significantly outperforms the ILP model for all sizes of use-cases, both in terms of computation time and quality of the obtained solution. Moreover, considering all 1000 use-cases, the heuristic saves up to 75\% of the computation time. This is done while 
sacrificing less than 1.5\% of the use-cases that it fails to solve and giving optimal results for the other 98.5\%}.

\vspace{3mm}

\subsubsection{Latency-dominated use-cases}

\begin{figure}[hbt]
\centering
\epsfig{file=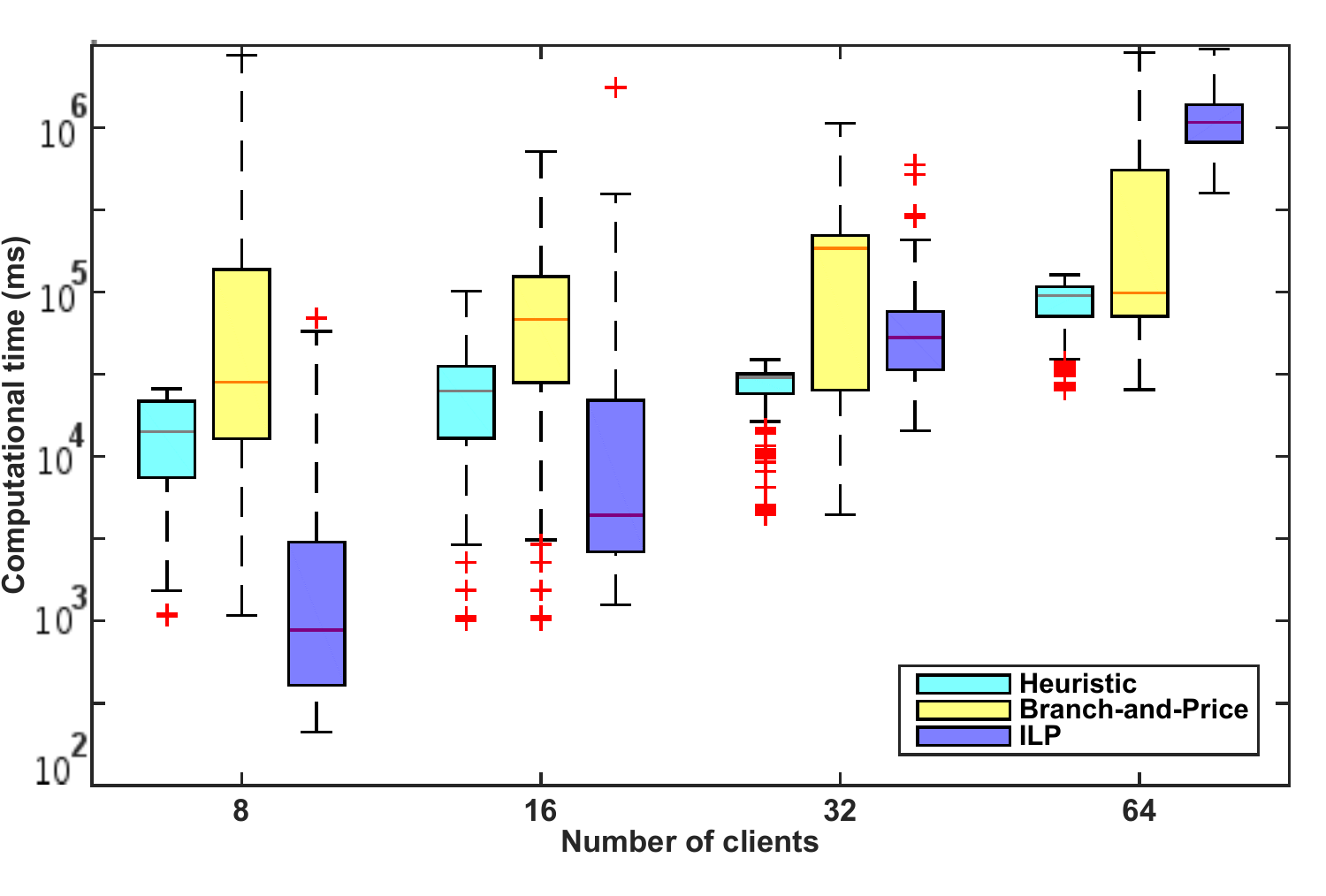,width=1\columnwidth}
\caption{Computation time distribution for the latency-dominated use-cases.}
\label{fig:timeLatDominated}
\end{figure}

The second group of use-cases that we focus on is latency-dominated use-cases. 
Since these use-cases are more complex than the bandwidth-dominated ones, instances with 128 clients take too long to run and are not included in the results.
The distribution of $log_{10}$ of the computation time is shown in Figure~\ref{fig:timeLatDominated}. 
Here, it is clear that for smaller use-cases with 8 and 16 clients, the ILP model significantly outperforms the branch-and-price approach. That is, the ILP runs in 2\% (15 minutes) and 42\% (176 minutes) of the branch-and-price computation time for 8 and 16 clients, respectively. 
However, with increasing number of clients the situation becomes different. For 64 clients, branch-and-price requires less time (total 26 hours versus the ILP with 65 hours, i.e. 2.5 times faster), but it is not able to find any feasible solution within the time limit for 5 use-cases. For the use-cases with 128 clients, running the heuristic 8 times in the beginning takes more than 50 minutes on average and in case the heuristic is not able to find a feasible solution, which happened in 14 use-cases out of 56, branch-and-price was able to find a solution in 2 use-cases out of 14 within the given time limit. 

As expected, the heuristic does not show good results on this type of use-cases with very demanding service latency requirements. Its failure rate goes up to 38\% for 16 clients, while the lowest failure rate is observed for 8 clients (23\%). However, the average distance from the best found solution does not exceed 1\% for any feasible use-case. It manages to save 40\%, 60\% and 90\% of the computation time of the least demanding optimal approach (whichever is faster, ILP or branch-and-price) for 16, 32 and 64 clients, respectively. 

The main reason the results for latency-dominated use-cases are different from their bandwidth-dominated counterparts is that latency-dominated 
clients require $O(\frameSize)$ instead of $O(\frameSize^{\:2})$ constraints. Therefore, the ILP model is able to quickly find a solution for larger problems and, as a consequence, branch-and-price starts to be faster only from 64 clients.

\emph{For the latency-dominated use-cases, we conclude that branch-and-price shows better results than the ILP starting from larger problem instances with 64 clients. The heuristic saves more time, sacrificing approximately the same number of solvable use-cases with increasing size of problem instances}.

\vspace{3mm}

\subsubsection{Mixed-dominated use-cases}

\begin{figure}[hbt]
\centering
\epsfig{file=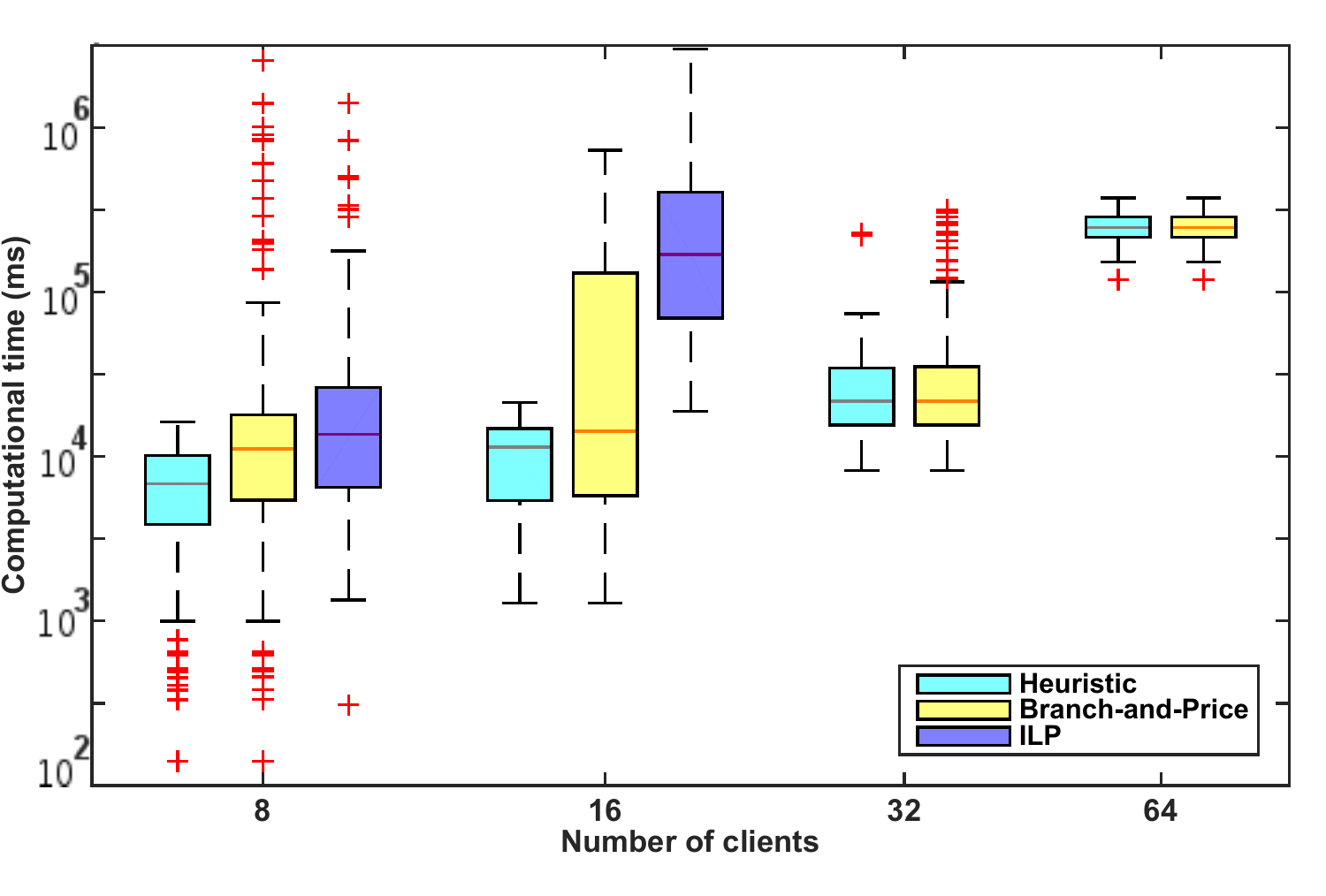,width=1\columnwidth}
\caption{Computation time distribution for the mixed-dominated use-cases.}
\label{fig:timeMixedDominated}
\end{figure}

Finally, experimental results for the set of mixed-dominated use-cases is shown in Figure~\ref{fig:timeMixedDominated}. 
Again, the ILP model fails to scale to use-cases with 32 clients, not even а feasible solution was found within a given time limit
in 172 out of 200 use-cases, which means a failure rate above 85\%. Thus, the results for 32 and 64 clients are only represented by the heuristic on the left and branch-and-price on the right. For the use-cases with 8 clients, the ILP model outperforms branch-and-price in terms of total computation time, although branch-and-price is slightly better in terms of median of the computation time. However, branch-and-price is not able to prove the optimality within the time limit for 3 use-cases, while the ILP failed only once. For 16 clients, the ILP model runs in 20 hours, while the branch-and price approach needs less than 5 hours, saving 77\% of the computation time. For 32 and 64 clients, branch-and-price requires on average less than 1 and 5 minutes on average for one use-case, respectively, demonstrating improved scalability.

For the mixed-dominated use-cases, the heuristic fails to find a feasible solution in 23 and 26 use-cases for 8 and 16 clients, respectively, and saves 95\% and 88\% of the computation time of the fastest optimal approach. The heuristic shows good results, especially on the use-cases with 32 and 64 clients, where it is able to solve almost all of them in 90 minutes and 14 hours, respectively, leaving only the proof of optimality to the branch-and-price approach. This result is caused by having more latency-dominated clients in the sets with 8 and 16 clients than in the sets of 32 and 64 clients, which makes the work for the heuristic easier. For the use-cases with 128 clients, the average time of running the heuristic 8 times is 75 minutes, which already exceeds given time limit.

\emph{The results for the mixed-dominated use-cases show significant reduction of computation time of the branch-and-price approach compared to the ILP model, starting approximately from use-cases with 16 clients, while giving optimal solutions for all use-cases. The heuristic saves 32\% of computation time on average, sacrificing approximately 8\% of the solvable use-cases.}

From these experiments, we confirm the exponential complexity of the problem, although our implementation solves an instance with 64 clients and 512 slots in less than 8 minutes on average. Moreover, the ILP model is not able to solve use-cases with more than 16 clients for bandwidth-dominated and mixed-dominated use-cases and 32 clients for latency-dominated use-cases. Thus, the branch-and-price approach is better for more complex use-cases, while ILP typically shows better results for the use-cases with smaller number of clients. More specifically, the proposed branch-and-price approach improves scalability from 16 to 128 clients for bandwidth-dominated use-cases and from 16 to 64 clients for mixed-dominated use-cases, while latency-dominated use-cases remain unchanged at 64 clients. In total, the work in this article improves scalability with approximately a factor 8.

To further improve the scalability of the branch-and-price approach, it is necessary to apply additional advanced methods on the given problem. Alternatively, it is possible to use the proposed heuristic, which enables saving up to 95\% of the computation time with loss of maximally 38\% of feasible use-cases, being 1\% in distance from the optimal solution on average. In contrast, the commonly used continuous slot assignment algorithm~\cite{Goossens13CODES,Foroutan13DSD,Goossens13DATE,Vink08} failed to find \emph{any} feasible solution in all 3000 use-cases. This simple yet common heuristic is hence unable to cover any use-cases of reasonable complexity.

\section{Case Study}
\label{case_study}

\begin{figure}[htb]
\centering
\epsfig{file=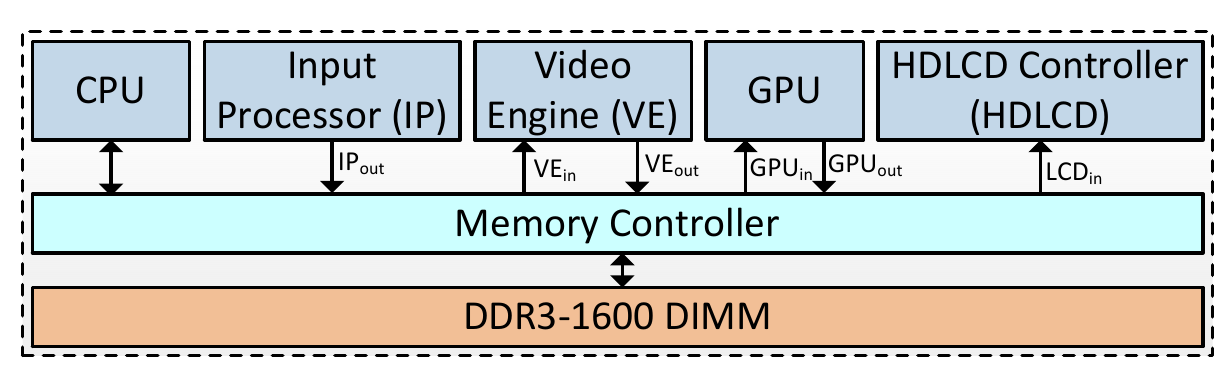,width=1.0\columnwidth}
\caption{Architecture of the HD video and graphics processing system.}
\label{fig:use_case}
\end{figure}

We now proceed by demonstrating the practical applicability of our proposed TDM configuration
methodology by applying it to a small case study of an HD video and
graphics processing system, where 7 memory clients share a 64-bit
DDR3-1600 memory DIMM~\cite{DDR3SPECf}. The considered system is illustrated in
Figure~\ref{fig:use_case}. Similarly to the multi-channel case study
in~\cite{gomony2015real}, we derive the client requirements from a
combination of the industrial systems in~\cite{Steffens08,Stevens2010}
and information about the memory traffic of the decoder application
from~\cite{Bonatto2011}. However, we assume 720p resolution instead of
1080p and that all memory requests have a fixed size of 128~B 
to be able to satisfy the requirements with a
single memory channel. 

The Input Processor receives an H.264 encoded YUV 4:2:0 video stream
with a resolution of $720 \times 480$, 12~bpp, at a frame rate of
25~fps~\cite{Steffens08}, and writes to memory (IP$_{out}$) at less
than 1~MB/s.  The Video Engine (VE) generates traffic by reading the
compressed video and reference frames for motion compensation
(VE$_{in}$), and writing decoder output (VE$_{out}$). The motion compensation
requires at least 285.1~MB/s to decode the video samples at a
resolution of $1280 \times 720$, 8~bpp, at
25~fps~\cite{Bonatto2011}. The bandwidth requirement to output the
decoded video image is 34.6~MB/s.  

The GPU is responsible for post-processing the decoded video.  The
bandwidth requirement depends on the complexity of the frame, but can
reach a peak bandwidth of 50~MB/frame in the
worst case~\cite{Stevens2010}. Its memory traffic can be split into
pixels read by the GPU for processing (GPU$_{in}$) and writing the frame
rendered by the GPU (GPU$_{out}$).  For GPU$_{in}$, we require a
guaranteed bandwidth of 1000~MB/s, which should be conservative given
that the peak bandwidth is not required continuously.  GPU$_{out}$
must communicate the complete uncompressed 720p video frame at 32~bpp
within the deadline of 40~ms (25~fps). With a burst size of 128~B,
this results in a maximum response time (finishing time - arrival
time) of 1388~ns per request. To provide a firm guarantee that all
data from this client arrives before the deadline, we separate this
into a service latency and a rate requirement according to the $\lr$
server approach. There are multiple $(\latency, \rho)$ pairs that can
satisfy a given response time requirement according to
Equation~\eqref{eq:finishing_time}, where a higher required bandwidth
results in a more relaxed service latency requirement.  Here, we
require a bandwidth of 184.3~MB/s, twice the continuous bandwidth that
is needed, to budget time for interference from other
clients. According to Equation~\eqref{eq:finishing_time}, this results
in a service latency requirement of 718~ns (574 clock cycles for an
800~MHz memory).

The HDLCD Controller (HDLCD) writes the image processed by the GPU to
the screen.  It is latency critical~\cite{Steffens08} and has a firm
deadline to ensure that data arrives in the frame buffer before the
screen is refreshed. Similarly to GPU$_{out}$, HDLCD requires at least
184.3~MB/s to output a frame every 40~ms. 
Note that each rendered frame
is displayed twice by the HDLCD controller to achieve a screen
refresh rate of 50~Hz with a frame rate of 25~fps. 
Lastly, a host CPU and its associated Direct Memory Access (DMA)
controller also require memory access with a total bandwidth of 150~MB/s to perform system-dependent
activities~\cite{Stevens2010}.

The derived requirements of the memory clients in the case study are 
summarized in Table~\ref{table:use_case_generation}. We conclude
the section by explaining how to transform
the requirements into the abstract units of rate and service latency 
(in slots) used by our approach. The rate is determined by
dividing the bandwidth requirement of the client with the minimum
guaranteed bandwidth provided by the memory controller. The service
latency requirement in slots is computed by dividing the latency
requirement in clock cycles by the WCET of a memory request.
Given a request size of 128~B and assuming the
real-time memory controller in~\cite{Akesson11DATE}, the WCET of a
memory request to a DDR3-1600 is 46 clock cycles at 800~MHz and the
memory guarantees a minimum bandwidth of 2149~MB/s~\cite{Goossens15-tc}.
For simplicity, we ignore effects of refresh interference in the memory,
which may increase the total memory access time over a video frame with
up to 3.5\% for this memory. 
The total required bandwidth of the clients in the case study is 1839.3~MB/s.
This corresponds to 85.6\% of the guaranteed bandwidth of the memory controller,
suggesting a suitably high load. In this use-case, all clients are bandwidth dominated.

\begin{table}[htb]
\rowcolors{1}{}{gray!35}
\caption{Client requirements in case study}
\label{table:use_case_generation} 
\centering
\begin{tabular}{|l|c|c|c|c|}
\hline
\emph{Client} & \emph{Bandwidth [MB/s]} & \emph{Latency [cc]} & $\requiredBandwidth$ & $\requiredLatency [slots]$ \\
\hline 
IP\textsubscript{out}  & 1.0    & -   & 0.0005 & -  \\ 
VE\textsubscript{in}   & 285.1  & -   & 0.1326 & -  \\
VE\textsubscript{out}  & 34.6   & -   & 0.0161 & - \\
GPU\textsubscript{in}  & 1000.0 & -   & 0.4652 & - \\
GPU\textsubscript{out} & 184.3  & 574 & 0.0858 & 12.5\\
LCD\textsubscript{in}  & 184.3  & 574 & 0.0858 & 12.5 \\ 
CPU                    & 150.0  & -   & 0.0698 & - \\ 
\hline
Total                  & 1839.3 &     & 0.8558 &\\
\hline
\end{tabular}
\end{table}

We apply our configuration methodologies to find the optimal TDM
schedule to satisfy the client requirements, while minimizing the
total allocated bandwidth. The frame size is set to $64$, which ensures 
that the use-case is solvable and provides a reasonable trade-off
between access granularity and total TDM schedule size for the number of
clients in the case study. Although the size of the problem is rather small, 
the branch-and-price approach results in more than 10 times reduction 
of the computation time compared to the ILP model. More specifically, the 
branch-and-price approach requires 138 milliseconds, while the ILP model
finishes in requires 1395 milliseconds. The reason branch-and-price is faster
even though the case study is small is that the heuristic provides an optimal
solution, as it typically does for bandwidth-dominated use-cases, significantly
reducing the computation time of the approach. From this case study, 
we conclude that \emph{branch-and-price can be advantageous not only for 
larger models, but also for the models of smaller sizes}. 

\section{Conclusions}
\label{conclusions}
This article introduces an approach that improves scalability of the existing approaches to configure resources shared by Time-Division Multiplexing (TDM) to satisfy bandwidth and latency requirements of real-time clients, while minimizing their total allocated rate to improve average performance of non-real-time clients. The problem considered here is to assign the slots to the clients, i.e. to find a TDM schedule with a given length. We propose an optimal approach that takes an existing integer linear programming model addressing the TDM configuration problem and wraps it in a branch-and-price framework to improve its scalability. In addition, a stand-alone heuristic that quickly finds a schedule is proposed for cases where an optimal solution is not required.

We experimentally evaluate the scalability of the branch-and-price approach and quantify the trade-off between computation time and solution quality for the proposed and existing optimal and heuristic algorithms. The results show that the branch-and-price approach can configure use-cases with up to 8x more clients, depending on the type of use-case. Average improvements are
in the range of 3-4x. The heuristic provides near-optimal solutions in 86\% of the use-cases with an average allocated bandwidth less than 0.26\% from the optimum in less than 50\% of the time of the fastest optimal approach (either ILP or branch-and-price). Throughout the experiments, our approach outperforms the ILP model on larger use-cases, while the ILP model shows better results on some use-cases of modest size. We also demonstrate the practical relevance of our approach by applying it to a case study
of a HD video and graphics processing system.

\section*{Acknowledgments}

This work was partially supported by the Horizon 2020 Programme of the European Commission under the Project HERCULES 688860 and Eaton European Innovation Centre.  




\bibliographystyle{template/IEEEtran}
\bibliography{template/IEEEabrv,bibliography}

\begin{thebibliography}{10}
\providecommand{\url}[1]{#1}
\csname url@samestyle\endcsname
\providecommand{\newblock}{\relax}
\providecommand{\bibinfo}[2]{#2}
\providecommand{\BIBentrySTDinterwordspacing}{\spaceskip=0pt\relax}
\providecommand{\BIBentryALTinterwordstretchfactor}{4}
\providecommand{\BIBentryALTinterwordspacing}{\spaceskip=\fontdimen2\font plus
\BIBentryALTinterwordstretchfactor\fontdimen3\font minus
  \fontdimen4\font\relax}
\providecommand{\BIBforeignlanguage}[2]{{%
\expandafter\ifx\csname l@#1\endcsname\relax
\typeout{** WARNING: IEEEtran.bst: No hyphenation pattern has been}%
\typeout{** loaded for the language `#1'. Using the pattern for}%
\typeout{** the default language instead.}%
\else
\language=\csname l@#1\endcsname
\fi
#2}}
\providecommand{\BIBdecl}{\relax}
\BIBdecl

\bibitem{ITRS-2011}
``{International Technology Roadmap for Semiconductors (ITRS)},'' 2011.

\bibitem{vanderWolf11}
P.~Van Der~Wolf and J.~Geuzebroek, ``{SoC infrastructures for predictable
  system integration},'' in \emph{Design, Automation \& Test in Europe
  Conference \& Exhibition (DATE), 2011}.\hskip 1em plus 0.5em minus
  0.4em\relax IEEE, 2011, pp. 1--6.

\bibitem{Kollig09}
P.~Kollig, C.~Osborne, and T.~Henriksson, ``Heterogeneous multi-core platform
  for consumer multimedia applications,'' in \emph{Proceedings of the
  Conference on Design, Automation and Test in Europe}, 2009, pp. 1254--1259.

\bibitem{Berkel09}
C.~Van~Berkel, ``{Multi-core for mobile phones},'' in \emph{Proceedings of the
  Conference on Design, Automation and Test in Europe}, 2009, pp. 1260--1265.

\bibitem{Feilet10}
D.~Feillet, ``A tutorial on column generation and branch-and-price for vehicle
  routing problems,'' \emph{4OR}, vol.~8, no.~4, pp. 407--424, 2010.

\bibitem{BandP_TDM}
A.~Minaeva, P.~Sucha, and B.~Akesson, ``{BandP\_TDM},''
  https://github.com/CTU-IIG/BandP\_TDM, 2015.

\bibitem{Lukasie2012}
M.~Lukasiewycz, R.~Schneider, D.~Goswami, and S.~Chakraborty, ``Modular
  scheduling of distributed heterogeneous time-triggered automotive systems,''
  \emph{Design Automation Conference (ASP-DAC), 2012 17th Asia and South
  Pacific}, pp. 665--670, January 2012.

\bibitem{gomony2015real}
M.~D. Gomony, B.~Akesson, and K.~Goossens, ``A real-time multichannel memory
  controller and optimal mapping of memory clients to memory channels,''
  \emph{ACM Transactions on Embedded Computing Systems (TECS)}, vol.~14, no.~2,
  p.~25, 2015.

\bibitem{Yi2009}
Y.~Yi, W.~Han, X.~Zhao, A.~Erdogan, and T.~Arslan, ``An {ILP} formulation for
  task mapping and scheduling on multi-core architectures,'' \emph{Design,
  Automation Test in Europe Conference Exhibition, 2009. DATE '09}, pp. 33--38,
  April 2009.

\bibitem{Lin2012}
J.~Lin, A.~Gerstlauer, and B.~Evans, ``Communication-aware heterogeneous
  multiprocessor mapping for real-time streaming systems,'' \emph{Journal of
  Signal Processing Systems}, vol.~69, no.~3, pp. 279--291, 2012.

\bibitem{Hanzalek2010}
Z.~Hanzalek, P.~Burget, and P.~Sucha, ``Profinet {IO IRT} message scheduling
  with temporal constraints,'' \emph{Industrial Informatics, IEEE Transactions
  on}, vol.~6, no.~3, pp. 369--380, Aug 2010.

\bibitem{Wildermann:2014:MDR}
S.~Wildermann, M.~Gla{\ss}, and J.~Teich, ``Multi-objective distributed
  run-time resource management for many-cores,'' in \emph{Proceedings of the
  conference on Design, Automation \& Test in Europe}, 2014, pp. 221:1--221:6.

\bibitem{Liu2008}
W.~Liu, M.~Yuan, X.~He, Z.~Gu, and X.~Liu, ``Efficient {SAT}-based mapping and
  scheduling of homogeneous synchronous dataflow graphs for throughput
  optimization,'' \emph{Real-Time Systems Symposium, 2008}, pp. 492--504, Nov
  2008.

\bibitem{Reimann:2011:SSS:2024724.2024817}
F.~Reimann, M.~Lukasiewycz, M.~Glass, C.~Haubelt, and J.~Teich, ``Symbolic
  system synthesis in the presence of stringent real-time constraints,'' in
  \emph{Proceedings of the 48th Design Automation Conference}, ser. DAC
  '11.\hskip 1em plus 0.5em minus 0.4em\relax New York, NY, USA: ACM, 2011, pp.
  393--398.

\bibitem{lukasiewycz12}
M.~Lukasiewycz and S.~Chakraborty, ``Concurrent architecture and schedule
  optimization of time-triggered automotive systems,'' in \emph{Proceedings of
  the eighth IEEE/ACM/IFIP international conference on Hardware/software
  codesign and system synthesis}.\hskip 1em plus 0.5em minus 0.4em\relax ACM,
  2012, pp. 383--392.

\bibitem{tamas2012synthesis}
D.~Tamas-Selicean, P.~Pop, and W.~Steiner, ``Synthesis of communication
  schedules for {TTEthernet}-based mixed-criticality systems,'' in
  \emph{Proceedings of the eighth IEEE/ACM/IFIP international conference on
  Hardware/software codesign and system synthesis}.\hskip 1em plus 0.5em minus
  0.4em\relax ACM, 2012, pp. 473--482.

\bibitem{Hansson07UNI}
A.~Hansson, K.~Goossens, and A.~R{\u{a}}dulescu, ``A unified approach to
  mapping and routing on a network-on-chip for both best-effort and guaranteed
  service traffic,'' \emph{VLSI design}, 2007.

\bibitem{Lu_slotallocation}
Z.~Lu and A.~Jantsch, ``Slot allocation using logical networks for {TDM}
  virtual-circuit configuration for network-on-chip,'' in \emph{Computer-Aided
  Design, 2007. ICCAD 2007. IEEE/ACM International Conference on}, 2007, pp.
  18--25.

\bibitem{hassan2015framework}
M.~Hassan, H.~Patel, and R.~Pellizzoni, ``A framework for scheduling {DRAM}
  memory accesses for multi-core mixed-time critical systems,'' \emph{Real-Time
  and Embedded Technology and Applications Symposium (RTAS), 2015 IEEE}, pp.
  307--316, 2015.

\bibitem{Goossens13CODES}
S.~Goossens, J.~Kuijsten, B.~Akesson, and K.~Goossens, ``A reconfigurable
  real-time {SDRAM} controller for mixed time-criticality systems,'' in
  \emph{Hardware/Software Codesign and System Synthesis (CODES+ ISSS), 2013
  International Conference on}.\hskip 1em plus 0.5em minus 0.4em\relax IEEE,
  2013.

\bibitem{Foroutan13DSD}
S.~Foroutan, B.~Akesson, K.~Goossens, and F.~Petrot, ``A general framework for
  average-case performance analysis of shared resources,'' in \emph{Digital
  System Design (DSD), 2013 Euromicro Conference on}.\hskip 1em plus 0.5em
  minus 0.4em\relax IEEE, 2013, pp. 78--85.

\bibitem{Goossens13DATE}
S.~Goossens, B.~Akesson, and K.~Goossens, ``Conservative open-page policy for
  mixed time-criticality memory controllers,'' in \emph{Proceedings of the
  Conference on Design, Automation and Test in Europe}.\hskip 1em plus 0.5em
  minus 0.4em\relax EDA Consortium, 2013, pp. 525--530.

\bibitem{Vink08}
J.~P. Vink, K.~van Berkel, and P.~van~der Wolf, ``Performance analysis of {SoC}
  architectures based on latency-rate servers,'' \emph{Design, Automation and
  Test in Europe, 2008. DATE'08}, pp. 200--205, 2008.

\bibitem{Akesson15}
B.~Akesson, A.~Minaeva, P.~{\v S}{\r u}cha, A.~Nelson, and Z.~Hanz{\'a}lek,
  ``Efficient configuration methodologies for time-division multiplexed
  resources,'' \emph{{Real-Time and Embedded Technology and Applications
  Symposium (RTAS)}}, pp. 161--171, 2015.

\bibitem{schenkelaars2011optimal}
T.~Schenkelaars, B.~Vermeulen, and K.~Goossens, ``Optimal scheduling of
  switched {FlexRay} networks,'' \emph{Design, Automation \& Test in Europe
  Conference \& Exhibition (DATE), 2011}, pp. 1--6, 2011.

\bibitem{Stiliadis98}
D.~Stiliadis and A.~Varma, ``Latency-rate servers: a general model for analysis
  of traffic scheduling algorithms,'' \emph{IEEE/ACM Transactions on Networking
  (ToN)}, vol.~6, no.~5, pp. 611--624, 1998.

\bibitem{Akesson08RTCSA}
B.~Akesson, L.~Steffens, E.~Strooisma, and K.~Goossens, ``Real-time scheduling
  using credit-controlled static-priority arbitration,'' in \emph{Embedded and
  Real-Time Computing Systems and Applications, 2008. RTCSA'08. 14th IEEE
  International Conference on}, 2008, pp. 3--14.

\bibitem{akesson11-book}
B.~Akesson and K.~Goossens, \emph{Memory Controllers for Real-Time Embedded
  Systems}, first edition~ed., ser. Embedded Systems Series.\hskip 1em plus
  0.5em minus 0.4em\relax Springer, 2011.

\bibitem{Wiggers07SCOPES}
M.~H. Wiggers, M.~J. Bekooij, and G.~J. Smit, ``Modelling run-time arbitration
  by latency-rate servers in dataflow graphs,'' in \emph{Proceedingsof the 10th
  international workshop on Software \& compilers for embedded systems}.\hskip
  1em plus 0.5em minus 0.4em\relax ACM, 2007, pp. 11--22.

\bibitem{Shah13DATE}
H.~Shah, A.~Knoll, and B.~Akesson, ``Bounding {SDRAM} interference: Detailed
  analysis vs. latency-rate analysis,'' in \emph{Design, Automation \& Test in
  Europe Conference \& Exhibition (DATE), 2013}.\hskip 1em plus 0.5em minus
  0.4em\relax IEEE, 2013, pp. 308--313.

\bibitem{Rodrigues15SOCP}
V.~Rodrigues, B.~Akesson, M.~Florido, S.~Melo~de Sousa, J.~P. Pedroso, and
  P.~Vasconcelos, ``Certifying execution time in multicores,'' 2015.

\bibitem{Sriram00}
S.~Sriram and S.~Bhattacharyya, \emph{{Embedded Multiprocessors: Scheduling and
  Synchronization}}.\hskip 1em plus 0.5em minus 0.4em\relax CRC, 2000.

\bibitem{Nelson2015}
A.~Nelson, K.~Goossens, and B.~Akesson, ``Dataflow formalisation of real-time
  streaming applications on a composable and predictable multi-processor
  {SoC},'' \emph{Journal of Systems Architecture}, 2015.

\bibitem{bhattacharyya1999synthesis}
S.~S. Bhattacharyya, P.~K. Murthy, and E.~A. Lee, ``Synthesis of embedded
  software from synchronous dataflow specifications,'' \emph{Journal of VLSI
  signal processing systems for signal, image and video technology}, vol.~21,
  no.~2, pp. 151--166, 1999.

\bibitem{stuijk2008throughput}
S.~Stuijk, M.~Geilen, and T.~Basten, ``Throughput-buffering trade-off
  exploration for cyclo-static and synchronous dataflow graphs,''
  \emph{Computers, IEEE Transactions on}, vol.~57, no.~10, pp. 1331--1345,
  2008.

\bibitem{Moreira07}
O.~Moreira, F.~Valente, and M.~Bekooij, ``Scheduling multiple independent
  hard-real-time jobs on a heterogeneous multiprocessor,'' in \emph{Proceedings
  of the 7th ACM \& IEEE international conference on Embedded software}.\hskip
  1em plus 0.5em minus 0.4em\relax ACM, 2007, pp. 57--66.

\bibitem{gomony2015generic}
M.~D. Gomony, J.~Garside, B.~Akesson, N.~Audsley, and K.~Goossens, ``A generic,
  scalable and globally arbitrated memory tree for shared {DRAM} access in
  real-time systems,'' in \emph{Proceedings of the 2015 Design, Automation \&
  Test in Europe Conference \& Exhibition}.\hskip 1em plus 0.5em minus
  0.4em\relax EDA Consortium, 2015, pp. 193--198.

\bibitem{bar2002minimizing}
A.~Bar-Noy, R.~Bhatia, J.~Naor, and B.~Schieber, ``Minimizing service and
  operation costs of periodic scheduling,'' \emph{Mathematics of Operations
  Research}, vol.~27, no.~3, pp. 518--544, 2002.

\bibitem{Vanderbei96}
R.~J. Vanderbei, ``Linear programming: foundations and extensions,'' 1996.

\bibitem{cplex}
{ILOG AMPL CPLEX}, ``{ System Version 7.0 User's Guide},'' \emph{ILOG CPLEX
  Division, Incline Village, NV}, 2000.

\bibitem{Huisman03}
D.~Huisman, R.~Jans, M.~Peeters, and A.~Wagelmans, ``Combining generation and
  lagrangian relaxation,'' \emph{ERIM Report Series Research in Management},
  2003.

\bibitem{DDR3SPECf}
\emph{{DDR3} {SDRAM} Specification}, {JESD79-3F}~ed., JEDEC Solid State
  Technology Association, 2012.

\bibitem{Steffens08}
L.~Steffens, M.~Agarwal, and P.~Wolf, ``Real-time analysis for memory access in
  media processing {SoCs}: A practical approach,'' in \emph{Real-Time Systems,
  2008. ECRTS'08. Euromicro Conference on}.\hskip 1em plus 0.5em minus
  0.4em\relax IEEE, 2008, pp. 255--265.

\bibitem{Stevens2010}
A.~Stevens, ``{QoS} for high-performance and power-efficient {HD} multimedia,''
  \emph{ARM White paper, http://wwww.arm.com}, 2010.

\bibitem{Bonatto2011}
A.~Bonatto, A.~Soares, and A.~Susin, ``Multichannel {SDRAM} controller design
  for {H.264/AVC} video decoder,'' \emph{In Programmable Logic (SPL), 2011 VII
  Southern Conference on}, pp. 137--142, 2011.

\bibitem{Akesson11DATE}
B.~Akesson and K.~Goossens, ``Architectures and modeling of predictable memory
  controllers for improved system integration,'' in \emph{Design, Automation \&
  Test in Europe Conference \& Exhibition (DATE), 2011}.\hskip 1em plus 0.5em
  minus 0.4em\relax IEEE, 2011, pp. 1--6.

\bibitem{Goossens15-tc}
S.~Goossens, K.~Chandrasekar, B.~Akesson, and K.~Goossens, ``Power/performance
  trade-offs in real-time {SDRAM} command scheduling,'' \emph{Computers, IEEE
  Transactions on}, vol.~PP, no.~99, pp. 1--1, 2015.

\end{thebibliography}

\newpage
\appendix

\section{Appendix}
\label{appendix}

This appendix shows a short illustration of the column generation algorithm described in Section~\ref{B&P}. The small problem instance used for this purpose considers 2 clients $\client_{1}, \client_{2}$ with $\requiredLatency = \left[3, 3\right]$ and $\requiredBandwidth = \left[0.5, 0.3\right]$. The TDM frame size is $\frameSize=10$. We assume initial columns with $\scheduleBP_{1,1}$, $\scheduleBP_{2,1}$ for clients $\client_{1}, \client_{2}$, respectively, according to:

\begin{eqnarray}
	\columnSet^R = & \left \{\scheduleBP_{1,1} =  \left[0, 0, 1, 1, 0, 0, 0, 1, 1, 1\right],\right. \\ &
	\left.\scheduleBP_{2,1} = \left[1, 1, 0, 0, 0, 1, 1, 0, 0, 0\right] \right\},
\end{eqnarray}
the restricted master model $MM(\columnSet^R)$ is formulated as:

\begin{align*}
	& Minimize:\ \frac{5}{10} \cdot \assignment_{1,1} +\frac{4}{10} \cdot \assignment_{2,1} +10\sum_{j \in \frameSizeSet}{\y_j}\\
	& \emph{subject to}:  \\
	& \ \ \ \ \ \ \ \ \ \ \ \ \ \ 0 \cdot \assignment_{1,1} + 1 \cdot \assignment_{2,1} \leq 1 + \y_1 \\ 
	&  \ \ \ \ \ \ \ \ \ \ \ \ \ \  0 \cdot \assignment_{1,1} + 1 \cdot \assignment_{2,1} \leq 1 + \y_2 \\
	& \ \ \ \ \ \ \ \ \ \ \ \ \ \ 1 \cdot \assignment_{1,1} + 0 \cdot \assignment_{2,1} \leq 1 + \y_3 \\
	& \ \ \ \ \ \ \ \ \ \ \ \ \ \ 1 \cdot \assignment_{1,1} + 0 \cdot \assignment_{2,1} \leq 1 + \y_4 \\
	& \ \ \ \ \ \ \ \ \ \ \ \ \ \ 0 \cdot \assignment_{1,1} + 0 \cdot \assignment_{2,1} \leq 1 + \y_5 \\
	& \ \ \ \ \ \ \ \ \ \ \ \ \ \ 0 \cdot \assignment_{1,1} + 1 \cdot \assignment_{2,1} \leq 1 + \y_6 \\
	& \ \ \ \ \ \ \ \ \ \ \ \ \ \ 0 \cdot \assignment_{1,1} + 1 \cdot \assignment_{2,1} \leq 1 + \y_7 \\
	& \ \ \ \ \ \ \ \ \ \ \ \ \ \ 1 \cdot \assignment_{1,1} + 0 \cdot \assignment_{2,1} \leq 1 + \y_8 \\
	& \ \ \ \ \ \ \ \ \ \ \ \ \ \ 1 \cdot \assignment_{1,1} + 0 \cdot \assignment_{2,1} \leq 1 + \y_9 \\
	& \ \ \ \ \ \ \ \ \ \ \ \ \ \ 1 \cdot \assignment_{1,1} + 0 \cdot \assignment_{2,1} \leq 1 + \y_{10} \\
	& \ \ \ \ \ \ \ \ \ \ \ \ \ \ 1 \cdot \assignment_{1,1} + 0 \cdot \assignment_{2,1} \geq 1 \nonumber \\
	& \ \ \ \ \ \ \ \ \ \ \ \ \ \ 0 \cdot \assignment_{1,1} + 1 \cdot \assignment_{2,1} \geq 1. \nonumber
\end{align*}

The trivial solution to the relaxed $MM(\columnSet^R)$ is $ \assignment_{1,1} = 1$, $ \assignment_{2,1} = 1$, $\y_j = 0~\forall j \in \frameSizeSet$ having objective function $\Criterium = 0.9$. The corresponding dual solution is $\dualsOne_j = 0~\forall j \in \frameSizeSet$, $\dualsTwo_1 = -0.5$, $\dualsTwo_2 = -0.4$.

When we formulate the sub-model from Equations~\eqref{BP8_1}-\eqref{BP11} for client $\client_{1}$ to obtain a new column, its solution is $\left[0, 1, 1, 0, 1, 1, 0, 0, 1, 0\right]$. However, its reduced price $\critSM =  \sum_{j \in \frameSizeSet} \scheduleMod_{1}^{j} \cdot 0 + 0.5 - 0.5 = 0$ (see Equation~\eqref{BP8_1}) is not negative, which means that this column cannot improve the objective function of $MM(\columnSet^R)$. On the other hand, the sub-model for client $\client_{2}$ finds $\left[1, 0, 0, 0, 1, 0, 0, 1, 0, 0\right]$ with reduced price $\sum_{j \in \frameSizeSet} \scheduleMod_{2}^{j} \cdot 0 + 0.3 - 0.4 = -0.1$, which has potential to improve the objective function of $MM(\columnSet^R)$.

When the new column for $\client_{2}$ is added into $MM(\columnSet^R)$ (see Step~7 in Figure~\ref{fig:BPScheme}),  $MM(\columnSet^R)$ is solved again in Step~2. In this case, the primal solution stays practically the same (i.e. $\assignment_{1,1} = 1$, $\assignment_{2,1} = 1$, $\assignment_{2,2} = 0$, $\y_j = 0~\forall j \in \frameSizeSet$), but the dual solution changes to $\dualsOne = \left[0, 0, 0, 0, 0, 0, 0, 0.1, 0, 0\right]$, $\dualsTwo_1 = -0.6$, $\dualsTwo_2 = -0.4$. For this dual solution, the sub-model for client $\client_{1}$ finds column $\left[0, 1, 1, 0, 1, 1, 0, 0, 1, 0\right]$ with reduced price $\critSM = 0 \cdot 0.1 + 0.5 - 0.6= -0.1$. This new column allows $MM(\columnSet^R)$ to find solution $\assignment_{1,1} = 0.5$, $\assignment_{2,1} = 0.5$, $\assignment_{2,2} = 0.5, \assignment_{1,2} = 0.5$, $\y_j = 0~\forall j \in F$ with a better value of the objective function equal to $\Criterium = 0.85$.

The last column $\scheduleBP_{2,3} = \left[1, 0, 0, 1, 0, 0, 1, 0, 0, 0\right]$ is added in the next iteration when $MM(\columnSet^R)$ achieves the optimal solution of the relaxed master model, which equals to $\Criterium = 0.8$. After that there is no column with negative reduced price and  column generation stops with

\begin{eqnarray}
	\Omega^R =  & \left\{\scheduleBP_{1,1} =  \left[0, 0, 1, 1, 0, 0, 0, 1, 1, 1\right],\right. \\
	& \left.\scheduleBP_{2,1} = \left[1, 1, 0, 0, 0, 1, 1, 0, 0, 0\right],\right. \\
	& \left.\scheduleBP_{2,2} = \left[1, 0, 0, 0, 1, 0, 0, 1, 0, 0\right],\right. \\
	& \left.\scheduleBP_{1,2} = \left[0, 1, 1, 0, 1, 1, 0, 0, 1, 0\right],\right. \\
	& \left.\scheduleBP_{2,3} = \left[1, 0, 0, 1, 0, 0, 1, 0, 0, 0\right] \right\}.
\end{eqnarray}

If the solution is integer (i.e. $\assignment_{1,1} = 0$, $\assignment_{2,1} = 0$, $\assignment_{2,2} = 0, \assignment_{1,2} = 1, \assignment_{2,3} = 1$, $\y_j = 0~\forall j \in \columnSet$), a new better solution was found and the branch-and-price algorithm continues by Step 8. Otherwise, the algorithm goes to Step 6 where the branching takes place.

\end{document}